\documentclass[journal]{IEEEtran}

\usepackage{times}
\usepackage{parskip}
\usepackage{subfig}
\usepackage{psfrag}
\usepackage{wrapfig}
\usepackage[latin1]{inputenc}
\usepackage{amssymb}
\usepackage{epsfig}
\usepackage{graphicx,times} 
\usepackage{amsfonts,amsmath,amssymb,array}
\usepackage{tabularx}   
\usepackage{colortbl}
\usepackage{longtable}
\usepackage{booktabs}
\usepackage{url}
\usepackage{icomma}
\usepackage{pstricks}
\usepackage{longtable}
\usepackage{multirow}
\usepackage{color}
\usepackage{graphicx}
\usepackage{cite}
\usepackage{picinpar}
\usepackage{amsmath}
\usepackage{amssymb}
\usepackage{amsthm}
\usepackage{mathtools}
\usepackage{url}
\usepackage{flushend}
\usepackage[latin1]{inputenc}
\usepackage{colortbl}
\usepackage{soul}
\usepackage{multirow}
\usepackage{pifont}
\usepackage{color}
\usepackage{alltt}
\usepackage{breakurl}
\usepackage{epstopdf}
\usepackage{pbox}
\usepackage{bbm}
\usepackage{algorithm2e}
\RestyleAlgo{ruled}

\newtheorem{definition}{Definition}

\newtheorem{lemma}{Lemma}

\newtheorem{theorem}{Theorem}
\newtheorem{bemerk}{Remark}

\DeclareMathOperator*{\argmax}{arg\,max}

\begin{document}

\title{Distributed Stackelberg Strategies in State-based Potential Games for Autonomous Decentralized Learning Manufacturing Systems

\thanks{This pre-print was submitted to IEEE Transactions on Systems, Man, and Cybernetics: Systems on July 31, 2024.}

}

\author{
\IEEEauthorblockN{Steve Yuwono\IEEEauthorrefmark{1}, Dorothea Schwung\IEEEauthorrefmark{2}, Andreas Schwung\IEEEauthorrefmark{1}}
\\ \IEEEauthorblockA{\IEEEauthorrefmark{1}South Westphalia University of Applied Sciences, Soest, Germany
    \\\{yuwono.steve, schwung.andreas\}@fh-swf.de}
\\ \IEEEauthorblockA{\IEEEauthorrefmark{2} Hochschule D{\"u}sseldorf University of Applied Sciences, D{\"u}sseldorf, Germany
\\ dorothea.schwung@hs-duesseldorf.de}
}

\maketitle
\thispagestyle{empty}
\pagestyle{empty}

\begin{abstract}

This article describes a novel game structure for autonomously optimizing decentralized manufacturing systems with multi-objective optimization challenges, namely Distributed Stackelberg Strategies in State-Based Potential Games (DS2-SbPG). DS2-SbPG integrates potential games and Stackelberg games, which improves the cooperative trade-off capabilities of potential games and the multi-objective optimization handling by Stackelberg games. Notably, all training procedures remain conducted in a fully distributed manner. DS2-SbPG offers a promising solution to finding optimal trade-offs between objectives by eliminating the complexities of setting up combined objective optimization functions for individual players in self-learning domains, particularly in real-world industrial settings with diverse and numerous objectives between the sub-systems. We further prove that DS2-SbPG constitutes a dynamic potential game that results in corresponding converge guarantees. Experimental validation conducted on a laboratory-scale testbed highlights the efficacy of DS2-SbPG and its two variants, such as DS2-SbPG for single-leader-follower and Stack DS2-SbPG for multi-leader-follower. The results show significant reductions in power consumption and improvements in overall performance, which signals the potential of DS2-SbPG in real-world applications.

\end{abstract}
\begin{IEEEkeywords}
Smart manufacturing, game theory, Stackelberg games, energy optimization, multi-objective optimization, distributed learning
\end{IEEEkeywords}

\section{Introduction}\label{sec:intro}

In modern manufacturing systems, the integration of Artificial Intelligence (AI), Internet of Things (IoT), and Cyber-Physical Systems (CPS) technologies has revolutionized operational efficiency by enabling functions like fault tolerance~\cite{Fernandes20222}, self-optimization~\cite{Schwung2021}, and anomaly detection~\cite{Kharitonov2022}. Modern systems prioritize adaptability and flexibility to extend productivity, leading to modular production units with hardware and software modularity controlled by decentralized control systems~\cite{Trentesaux2009, Bakule2008}. Consequently, such evolution results in the necessity of developing distributed optimization methodologies, which assign systems to adjust to fluctuating demands within flexible manufacturing environments dynamically. Modular manufacturing systems are often considered distributed multi-agent systems (MAS)~\cite{Wooldridge2009}, where each sub-system has multiple distinct objectives. This poses challenges in determining the main focus of each agent and its local objective function to be aligned with the global system's objectives. This complexity becomes even more pronounced with a large number of agents with diverse and conflicting objectives, as found in large-scale manufacturing systems~\cite{Yuwono2023b}. 

Various machine learning methodologies have been explored for achieving distributed self-optimization in MAS, with multi-agent reinforcement learning (MARL)~\cite{Sutton2018} standing out due to its adaptability. MARL has found applications in routing solutions for mobile sensor networks~\cite{Okine2024}, robotics~\cite{Orr2023, Chen2017}, resource management for unmanned aerial vehicles~\cite{Cui2019, Peng2020}, and cyber security~\cite{Nguyen2021}. Despite its successes, recent implementations have revealed limitations in real-world scenarios. Firstly, most MARL approaches operate in a centralized instance~\cite{Lowe2017, Rashid2020, Foerster2018}, which leads to communication overhead and scalability issues for large-scale systems~\cite{Canese2021}. Secondly, the computational resources of MARL required are often beyond what is available in real industrial settings~\cite{Dulac2021}. Thirdly, MARL typically requires extensive data and training time to converge~\cite{Dulac2021}. Fourthly, MARL restricts dynamic interaction between agents due to its limitation to the Markov Decision Process (MDP) framework~\cite{Nowe2021}. Additionally, coordination problems between agents are not well-addressed in MARL. Lastly, managing diverse multi-objectives across subsystems leads to complexity in defining their reward functions, which requires careful consideration for optimal trade-offs.

In recent research, we addressed the limitations of MARL by introducing a dynamic state-based game theoretical (GT) approach integrating model-free~\cite{Schwung2020, Schwung2023} and model-based~\cite{Yuwono2023b, Yuwono2023c} learning principles. Our methodology follows cooperative potential games~\cite{Monderer1996, Marden2012, Zazo2016}, which enables agents to collaborate effectively while maintaining fully distributed optimization methods. State-based Potential Games (SbPGs)~\cite{Schwung2020} facilitate coordination among agents in self-learning modular production units which, consistently converge to Nash equilibrium and outperform MARL. However, challenges in handling diverse multi-objectives across subsystems remain unsolved. In~\cite{Schwung2020, Schwung2023, Yuwono2023b, Yuwono2023c}, we employed a weighted utility function to prioritize critical objectives but acknowledge this may not be the most optimized solution, where determining the weights and designing utility functions remains difficult, with no optimality guarantees. Automated parameter tuning through grid search~\cite{Bergstra2012} was used but was time-consuming and may not guarantee optimality. Our current study aims to develop a solution where each player manages objectives independently without concern for inter-objective weights while retaining the effectiveness of SbPGs.

GT includes various structures and strategies~\cite{Owen2013, Bauso2016}, including the Stackelberg strategy~\cite{Simaan1973}, characterized by a hierarchical order of play with a leader-and-follower dynamic. In Stackelberg games, decision-making unfolds sequentially, which allows for explicit modelling of strategic interactions' hierarchical nature and potentially more efficient outcomes compared to simultaneous-move games. These games implicitly prioritize objectives by designating leaders and followers, which aims for optimality at the Stackelberg equilibrium~\cite{Stackelberg2010, Fiez2020}. In our research, we propose a method to incorporate Stackelberg strategies in SbPGs within each player, which assigns roles based on different objectives to simplify defining utility functions for multi-objective optimizations. Our goal is to ensure convergence of the proposed approach while satisfying the principles of both potential games and Stackelberg equilibrium.

Here are the contributions of our paper:
\begin{itemize}
    \item We introduce a novel approach, namely distributed Stackelberg strategies in SbPG (DS2-SbPG), to address the complexities of multi-objective optimization in distributed settings.
    \item We present and analyze two DS2-SbPG variants, which are one with single-leader-follower dynamics and another, Stack DS2-SbPG, for scenarios with multiple leaders-followers. Additionally, we propose and validate an improved learning algorithm for optimizing the leader-follower dynamics in DS2-SbPG.
    \item We demonstrate their effectiveness by combining potential games and Stackelberg strategies while proving their convergence guarantees.
    \item We validate the practical utility through implementation in a laboratory test-scale system, enhancing overall system performance and reducing power consumption by 10.04\% by DS2-SbPG and 10.61\% by Stack DS2-SbPG compared to the standard SbPG.
\end{itemize}

This paper contains seven sections. Sec.~\ref{sec:review} provides a literature review. Sec.~\ref{sec:prob} describes the problem. In Sec.~\ref{sec:pg}, we discuss preliminary game structures. Sec.~\ref{sec:ds2sbpg} elaborates on the proposed DS2-SbPG framework, including its variants, learning algorithm, and proof of convergence. Sec.~\ref{sec:test} outlines the training setup and experimental findings. Finally, Sec.~\ref{sec:conc} summarizes the paper and annotations for future research.

\section{Literature Review}
\label{sec:review}

In this section, we review the literature on multi-objective optimizations, decentralized learning manufacturing systems, and dynamic GT with engineering applications.

\subsection{Multi-Objective Optimizations}
\label{sec:review_2}

Multi-objective optimization involves simultaneously optimizing multiple conflicting objectives, often represented mathematically as the minimization or maximization of multiple objective functions~\cite{Miettinen1999, Branke2008}. This field contains diverse mathematical models and algorithms aimed at finding trade-off solutions along the Pareto frontier~\cite{Ngatchou2005, Kalyanmoy2016}, where improving one objective results in the degradation of another. Its applications span various domains~\cite{Sharma2022}, e.g. finance, logistics, and resource allocation, where decision-makers must balance competing objectives. In modern control technology, state-of-the-art methods in multi-objective optimization employ advanced algorithms to efficiently explore the solution space and identify Pareto-optimal solutions~\cite{Ngatchou2005, Kalyanmoy2016}, like genetic algorithms, particle swarm optimization, and evolutionary strategies.

Another subset of multi-objective optimizations is constrained optimization~\cite{Bertsekas2014}, which introduces additional constraints alongside objective functions that require specialized algorithms for finding solutions, e.g. SAT (satisfiability) methods~\cite{Biere2009}. While multi-objective optimizations offer decision-makers a range of trade-off solutions, they also present challenges~\cite{Wenlan2019}, such as difficulty in determining objective weights and the absence of a single optimal solution that simultaneously optimizes all objectives. In the self-learning domain, multi-objective optimization is applied in defining objective functions guiding the learning process, for instance, reward functions in deep MARL~\cite{Mannion2018, Yuandou2019} and utility functions in dynamic GT~\cite{Schwung2023, Yuwono2023b}. Previous approaches often aggregated multiple objectives into a single objective function with weighted objectives, which poses challenges in accurately defining these weights~\cite{Wenlan2019}. In~\cite{Diehl2023}, a potential solution is introduced that utilises neural networks to learn the weights of optimization objectives, serving as an inductive bias. Meanwhile, in our research, we propose an approach that maintains each objective function independently during the learning process, which allows decision-makers to navigate trade-offs internally through Stackelberg strategies.

\subsection{Decentralized Learning Manufacturing Systems}
\label{sec:review_3}

Decentralized learning in manufacturing systems~\cite{Mourtzis2013, Jiewu2023} involves distributed knowledge acquisition and performance enhancement across multiple entities. This approach, often viewed as MAS~\cite{Wooldridge2009}, finds applications in production scheduling~\cite{Du2024}, resource allocation~\cite{Lee2024}, and predictive maintenance~\cite{Rokhforoz2021}. Decentralized learning in MAS enhances adaptability, resilience, and efficiency in complex manufacturing environments by enabling agents to learn and decide based on local information autonomously. This enables the system to respond effectively to dynamic changes and uncertainties. In self-learning domains, state-of-the-art approaches contain methodologies like MARL~\cite{Sutton2018}, dynamic GT~\cite{Owen2013, Bauso2016}, federated learning~\cite{Li2020}, and swarm intelligence~\cite{Ilie2013}. These methodologies enable agents in MAS to learn and adapt their strategies over time, which leads to improved system performance and productivity. Our prior research~\cite{Schwung2023, Yuwono2023b, Yuwono2023c} suggests that dynamic GT is more proficient and applicable in self-learning distributed MAS compared to MARL~\cite{Schwung2021} and model predictive controller~\cite{Yuwono2023a}, as it facilitates coordination among agents that boosts cooperative strategies to address competing and diverse multi-objectives. Therefore, this study opts for dynamic GT for decentralized self-learning in MAS.

\subsection{Dynamic GT with Engineering Applications}
\label{sec:review_1}

GT~\cite{Owen2013, Bauso2016} is a mathematical framework used to model and analyze interactions between rational decision-makers, known as players, who aim to maximize their utility or payoff. Various game structures in GT represent different scenarios and dynamics of strategic interactions, such as competition, cooperation, or coordination. Examples include potential games~\cite{Monderer1996}, Stackelberg games~\cite{Simaan1973}, Colonel Blotto games~\cite{Borel1953}, and congestion games~\cite{Rosenthal1973}. Furthermore, dynamic GT~\cite{Tamer2018} extends traditional GT to analyze situations where players' decisions evolve over time, which captures the sequential nature of actions and the feedback loop between decisions and outcomes. In engineering, dynamic GT finds various applications, for instance, in resource allocation~\cite{Lim2021}, edge computing on unmanned aerial vehicles~\cite{Zhaolong2021}, and optimization of manufacturing processes~\cite{Schwung2023}.

GT approaches are particularly valuable in distributed self-learning MAS, where agents interact and make adaptive decisions based on local information, as in~\cite{Schwung2023, Yuwono2023b, Yuwono2023c}. A notable advancement is the incorporation of state-based systems into game structures, which leads to SbPGs~\cite{Schwung2020}. In recent years, SbPGs have been augmented with advancements in model-based learning~\cite{Yuwono2023b, Yuwono2023c}. However, in multi-objective optimizations, determining the optimal combination of competing objectives remains challenging. Stackelberg games~\cite{Simaan1973}, with their hierarchical decision-making structure, aim for optimality at the Stackelberg equilibrium~\cite{Stackelberg2010, Fiez2020} and have proven effective in various engineering fields, like in power control communications~\cite{Yan2014} and security issues~\cite{Yuzhe2018, Luliang2018}. Hence, in this study, we propose improving SbPG by enabling each player to employ a Stackelberg strategy internally to achieve consensus on diverse conflicting multi-objectives.

\section{Problem Descriptions}
\label{sec:prob}

\begin{figure}[t]
	\centering
	\includegraphics[width=1.0\linewidth,keepaspectratio]{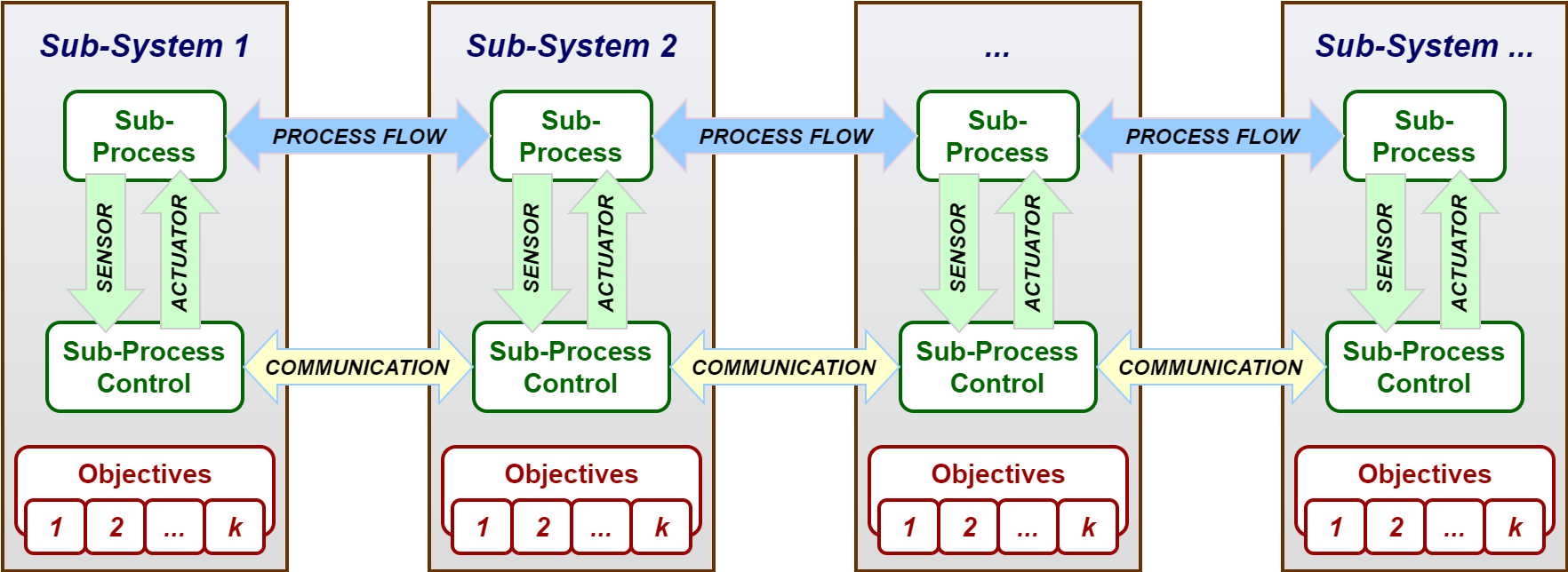}
	\caption{An illustration of modular production units configured with multiple objectives within each module.}
	\label{fig:prob_modular}
\end{figure}

This section provides the problem description in this study, such as autonomous learning within fully distributed manufacturing systems, where we focus on modular systems divided into several subsystems, each with its own local control system and diverse multi-objectives, as shown in Fig.~\ref{fig:prob_modular}. Each sub-system includes one or more actuators, whereas, in GT terms, a controllable actuator can be considered as a player $i$. Our primary goal is to achieve self-optimization of these systems in a fully distributed manner, which eliminates the necessity for a centralized control instance as well as allows for scalable, flexible, and generally reusable execution across various modules through instantiations.  



\begin{figure}[t]
	\centering
	\includegraphics[width=1.0\linewidth,keepaspectratio]{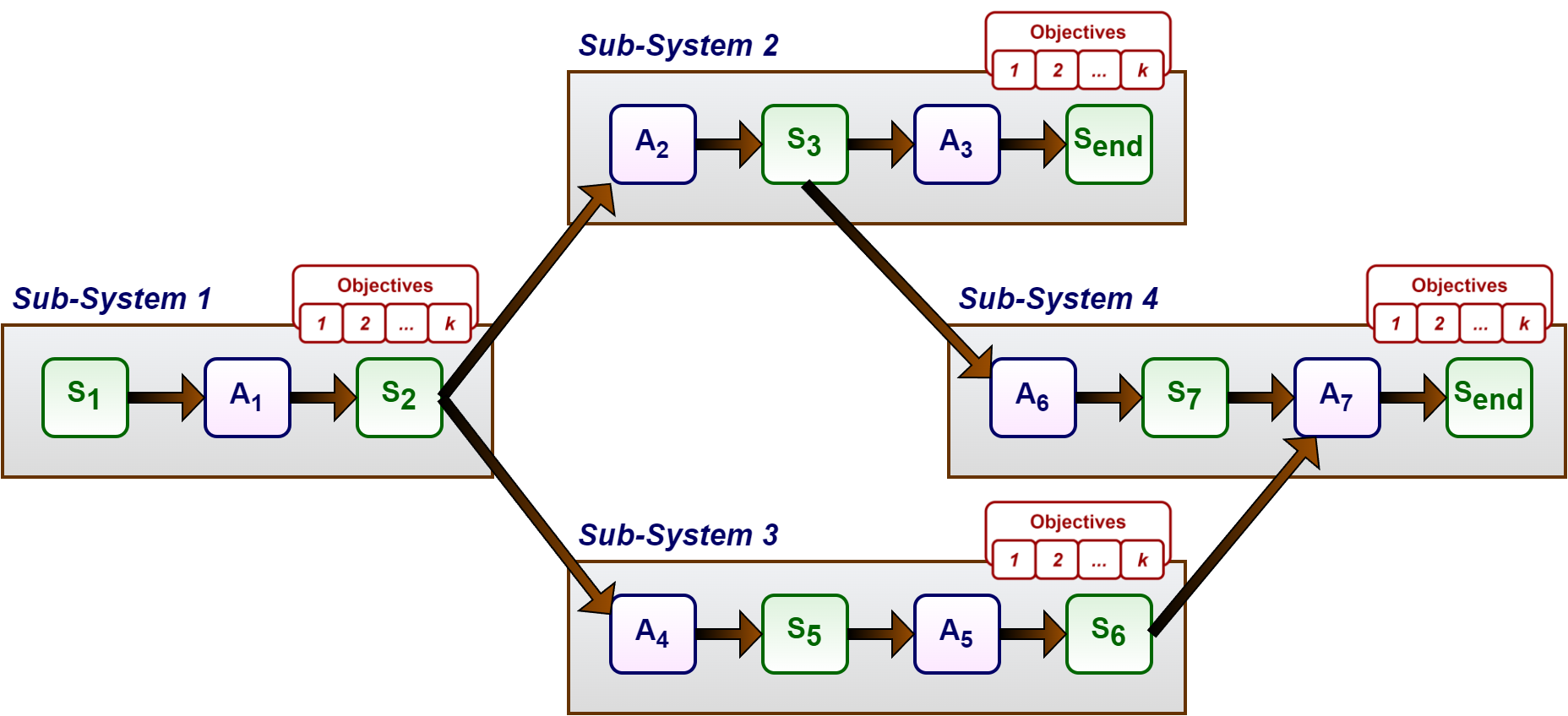}
	\caption{A schematic diagram of a production chain with serial-parallel connected subsystems, each with multiple objectives within its module.}
	\label{fig:prob_chain}
\end{figure}

The considered distributed system has been represented to a production chain in~\cite{Schwung2020} based on graph theory~\cite{Yamamoto2015}, where we model the production chain in both serial and serial-parallel process chains as an alternating sequence of actuators (e.g., motors, conveyors, feeders, pumps) and physical states representing the process status, as shown in Fig.~\ref{fig:prob_chain}. These actuators are expected to demonstrate a hybrid actuation system with both continuous and discrete operational behaviours. The production chain is defined as a dynamic sequence involving actuators $\mathcal{N}={1,\ldots,N}$ with sets of continuous or discrete action ${A}_i \subset \mathbb{R}^c \times \mathbb{N}^d$ and a group of states $\mathcal{S}\subset \mathbb{R}^m$. Then,  $\mathcal{E}$ denotes the edges,  excludes edges of the condition $e=(A_i,A_j)$ and $e=(s_i,s_j)$ with $A_i, A_j \in \mathcal{N}$ and $s_i, s_j \in \mathcal{S}$.  For each actuator $A_i\in \mathcal{N}$, two neighbouring states are introduced, including foregoing neighbor states $\mathcal{S}^{A_i}_{prior}=\{s_j \in \mathcal{S}|\exists e=(s_j,A_i)\in \mathcal{E} \}$ and following neighbor states $\mathcal{S}^{A_i}_{next}=\{s_j \in \mathcal{S}|\exists e=(A_i,s_j)\in \mathcal{E}\}$.

The assumption of the production chain consisting solely of sequences of states and actions is not restrictive. An arrangement involving more than two states can be reconstructed into a common state vector, and similar considerations apply to actions. Decentralized production lines are common in modern industrial environments. This distributed production scenario has applications across various fields in the process industry, including transportation processes of coal and grains, chemical plants, food production and pharmaceuticals.

We assume that each player $i$ has a local utility $U_i(a_i, S^{A_i})$ where ${S}^{A_i} \in \mathcal{S}^{A_i} = \mathcal{S}^{A_i}_{prior} \cup \mathcal{S}^{A_i}_{next} \cup \mathcal{S}^{g}$, $\mathcal{S}^{g}$ representing the states tied to global objectives. The local utility $U_i(a_i, S^{A_i})$ is assumed to consist of several, individual sub-objectives, $u_1^{(i)}, u_2^{(i)}, \ldots, u_k^{(i)}$, with each utility function representing a different aspect of the problem. In terms of production systems, this set of objectives for individual players $i$ includes e.g. satisfying production demand, minimizing power consumption, avoiding bottlenecks and satisfying production constraints. 

Consequently, the global objective is this work is to maximize the overall system utility $\phi$
\begin{equation} \label{Eq:overall_util}
\max_{a_i\in {A}_i} \phi({a},{S}),
\end{equation}
by jointly maximizing local utilities
\begin{equation} \label{Eq:multi_objective}
\max_{a_i\in {A}_i} ( u_1^{(i)}(s_i, a_i), u_2^{(i)}(s_i, a_i), \ldots, u_k^{(i)}(s_i, a_i)),
\end{equation}
where $k$ denotes the number of objective functions to be optimized.

In prior research~\cite{Schwung2020, Schwung2023, Yuwono2023b}, the multiple objectives are combined into a single function
\begin{equation} \label{Eq:util_sbpg}
U_i(s_i, a_i) = \omega_1 u_1^{(i)}(s_i, a_i) + \omega_2 u_2^{(i)}(s_i, a_i) + \ldots + \omega_k u_k^{(i)}(s_i, a_i), 
\end{equation}
where $a_i \in {A}_i$ and the weight parameter $\omega_k$ indicates the importance allocated to objective $k$. However, determining appropriate $\omega_k$ proves to be a non-trivial effort due to its high sensitivity to the direction of the learners, which makes it difficult to guarantee the optimal solutions for the controlled systems.

Furthermore, in most production environments, the local objective might have a different priority level, for instance, satisfying production demand is typically a primary objective, while reducing energy consumption serves as a secondary objective. Hence, the local objectives within each player $i$ must be suitably combined to maximize the global objective while considering the priority level. Achieving this alignment can be challenging due to potential conflicts among the local objectives and issues related to prioritization.

Two major challenges arise from the discussed problem descriptions. The first challenge is managing conflicting and diverse utility functions between players to optimize the global objectives. It is important to ensure that all players cooperate rather than act selfishly, a problem handled by SbPG~\cite{Schwung2020}, which has been proven to converge. The second challenge is managing the multi-objectives within each local objective, where the previously used weighted method cannot guarantee optimal solutions. In this study, we address prioritization among local objectives by employing leader-follower games, assigning objectives as either leader or follower to effectively manage their prioritization. We focus on the learning process within each player $i$ and employ Stackelberg strategies internally to manage trade-offs between objectives while maintaining SbPG as the primary game structure for interactions between players. This approach results in a combination of distributed Stackelberg strategies and SbPGs.

\section{Preliminary Game Structures}
\label{sec:pg}

This section focuses on two foundational game structures, namely SbPGs and Stackelberg games, which form the basis of our proposed game structure.

\subsection{State-based Potential Games}
\label{sec:pg_2}

Potential games model and analyze strategic interactions among rational agents where players' payoffs (utilities) depend on their actions and the environment's state. These interactions are evaluated using a scalar potential function $\phi$, a global objective function. Each player's utility function $U_i$ is impacted by the overall system's states rather than individual actions $a_i$. SbPGs~\cite{Marden2012} extend potential games~\cite{Monderer1996} by incorporating state information into strategic interactions.

In~\cite{Schwung2020}, the game is extended for self-optimizing modular production units and solving multi-objective optimization problems by incorporating the set of states $S$ and the state transition process $P$. Here is the formal definition of a SbPG~\cite{Schwung2020}:
\begin{definition}
    A game $\Gamma(\mathcal{N}, \mathcal{A}, \{U_i\}, {S}, P, \phi)$ constitutes an SbPG, when it ensures that the potential function $\phi:  a \times  {S} \rightarrow \mathcal{R}^{c_i}$ within the game and each pairing of action-state $[a,s]\in  {A}\times {S}$ meets the following requirements:
    \begin{align}\label{eq:potcondsbpg}
    U_i(a_i,s) - U_i({a}^{\prime}_i, {a}_{-i},s) = \phi(a_i,s) - \phi({a}^{\prime}_i, {a}_{-i},s),
    \end{align}
    and
    \begin{align}\label{eq:condsbpg}
    \phi(a_i,s^{\prime}) \geq \phi(a_i,s),
    \end{align}
    for any state $s^{\prime}$ in $P(a,s)$, where $\mathcal{R}^{c_i}$ characterises continuous actions.
\end{definition} 

A notable characteristic of (exact) potential games is their convergence to Nash equilibrium points under best-response dynamics~\cite{Marden2012}. In~\cite{Zazo2016}, the criteria of being an SbPG are demonstrated, where the convergence is sustained. This convergence is further proven for self-optimizing modular production units in~\cite{Schwung2020}. In this study, we maintain SbPGs as the primary game structure to facilitate cooperative interactions among players. However, SbPGs do not guarantee effective management of multi-objectives for each player individually. As a result, we integrate a strategy to manage these objectives distributively using a hierarchical Stackelberg structure.


\subsection{Stackelberg Games}
\label{sec:pg_1}

Stackelberg games~\cite{Simaan1973} are a GT framework that explores hierarchical interactions among rational players, unlike most GT scenarios where players act simultaneously. There are two roles in this game: leader and follower. The leader can pre-commit to a strategy, in which the follower subsequently responds to the leader's actions. This hierarchical form is often suitable for a non-cooperative game framework but can also be adapted for a cooperative game. Optimization methods such as dynamic programming, variational inequalities, or Stackelberg equilibrium concepts~\cite{Stackelberg2010} are commonly employed to analyze and find equilibrium solutions in Stackelberg games. Here is the formal definition of a Stackelberg game~\cite{Fiez2020}:
\begin{definition}\label{def:sg_1}
    We consider a cooperative Stackelberg game between two players where Player 1 is the leader with an objective $U_1:A_1\rightarrow\mathbb{R}$ and Player 2 is the follower with an objective $U_2:A_2\rightarrow\mathbb{R}$. Their action spaces are defined as $A=A_1\times A_2\in\mathbb{R}^{m}$ with $A_1\in\mathbb{R}^{m_1}$ and $A_2\in\mathbb{R}^{m_2}$. Thus, the leader and follower aim to solve the following optimization problem:
    \begin{equation}\label{eq:sg_1}
        \max_{{a_1}\in A_1} \{U_1({a_1},{a_2}) | a_2 \in \argmax_{{y}\in A_2}U_2({a_1},{y})\},
    \end{equation}
    \begin{equation}\label{eq:sg_2}
        \max_{{a_2}\in A_2} U_2({a_1}, {a_2}).
    \end{equation}
\end{definition}

\begin{bemerk}
    Definition~\ref{def:sg_1} highlights its contrast to simultaneous games such as in SbPGs, where each player $i$ faces the optimization problem $\max_{{a_i}\in A_i} U_i({a_i}, {a_{\_i}})$, leading to Eq.~\eqref{Eq:util_sbpg}.
\end{bemerk}

The sequential decision-making process in Stackelberg games establishes a hierarchical structure where the leader anticipates and influences the follower's actions. In most cases, the most critical objective within the set of multi-objectives assumes a higher hierarchical position, which acts as the leader objective. The final coalition decision $a_i$ is then derived by harmonizing the decisions of the leader and the follower objective. The leader aims to optimize its utility given the follower's response, while the follower plays the best response to the leader's actions. Hence, to manage multiple objectives without relying on weighted utility functions in SbPGs, we introduce the hierarchical approach of the Stackelberg strategy to each player.

\section{Distributed Stackelberg Strategies in State-Based Potential Games}
\label{sec:ds2sbpg}

\begin{figure}[t]
	\centering
	\includegraphics[width=1.0\linewidth,keepaspectratio]{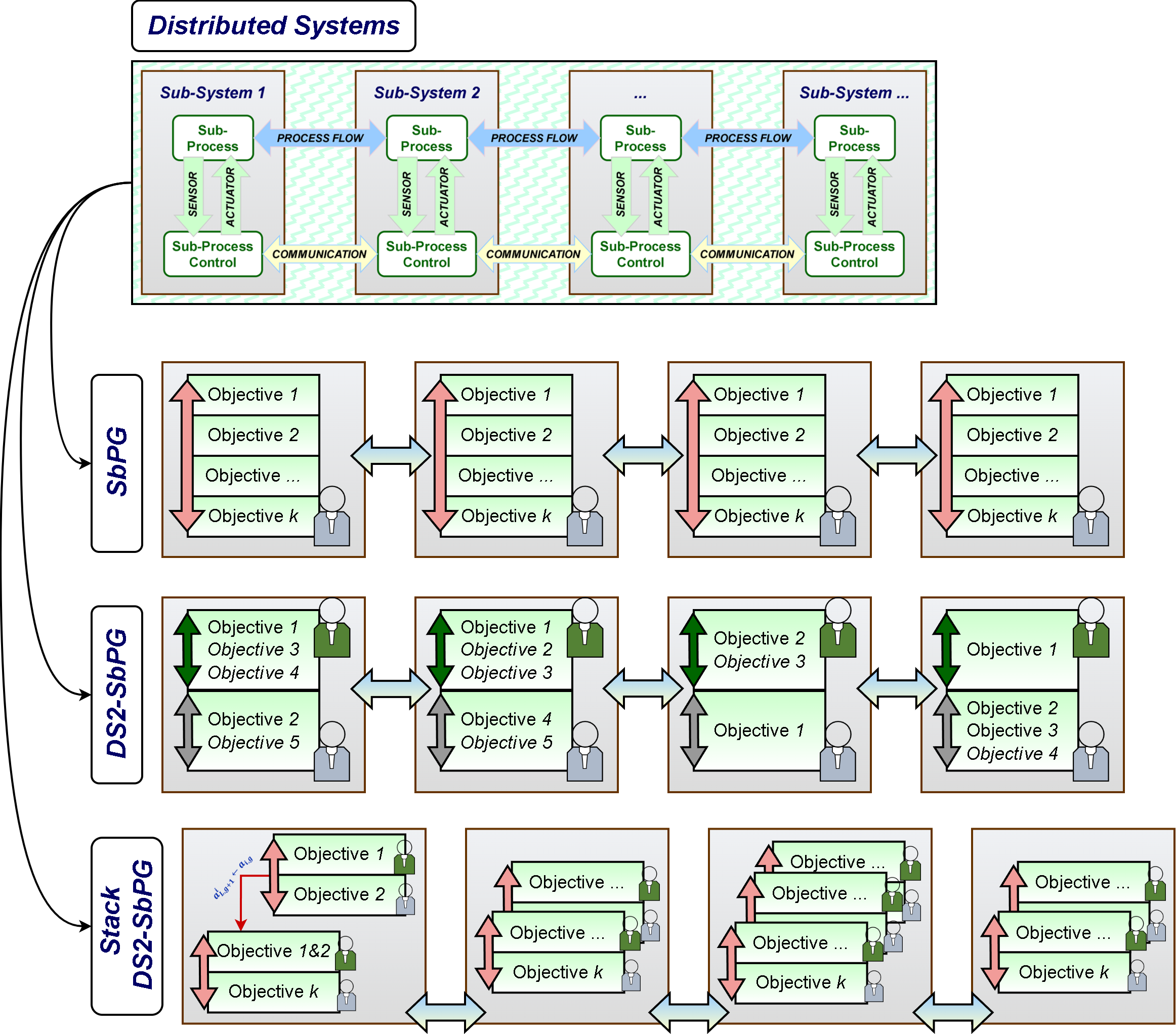}
	\caption{An overview between SbPG and two variants of DS2-SbPG in multi-objective optimizations for distributed manufacturing systems.}
	\label{fig:overview}
\end{figure}

This section introduces the fundamental and two variations of DS2-SbPG: (1) DS2-SbPG for single-leader-follower objective scenarios and (2) Stack DS2-SbPG for multi-leader-follower objective scenarios. We also detail the learning algorithm used to optimize the policies of leaders and followers, along with the training mechanism of DS2-SbPGs using this algorithm. Furthermore, we provide proof of convergence for the proposed DS2-SbPG and its variants. Fig.~\ref{fig:overview} provides an overview of SbPG and both DS2-SbPG variants in multi-objective optimization problems.

\subsection{DS2-SbPG for Single-Leader-Follower Objective}
\label{sec:ds2sbpg_1}

\begin{figure}[t]
	\centering
	\includegraphics[width=1.0\linewidth,keepaspectratio]{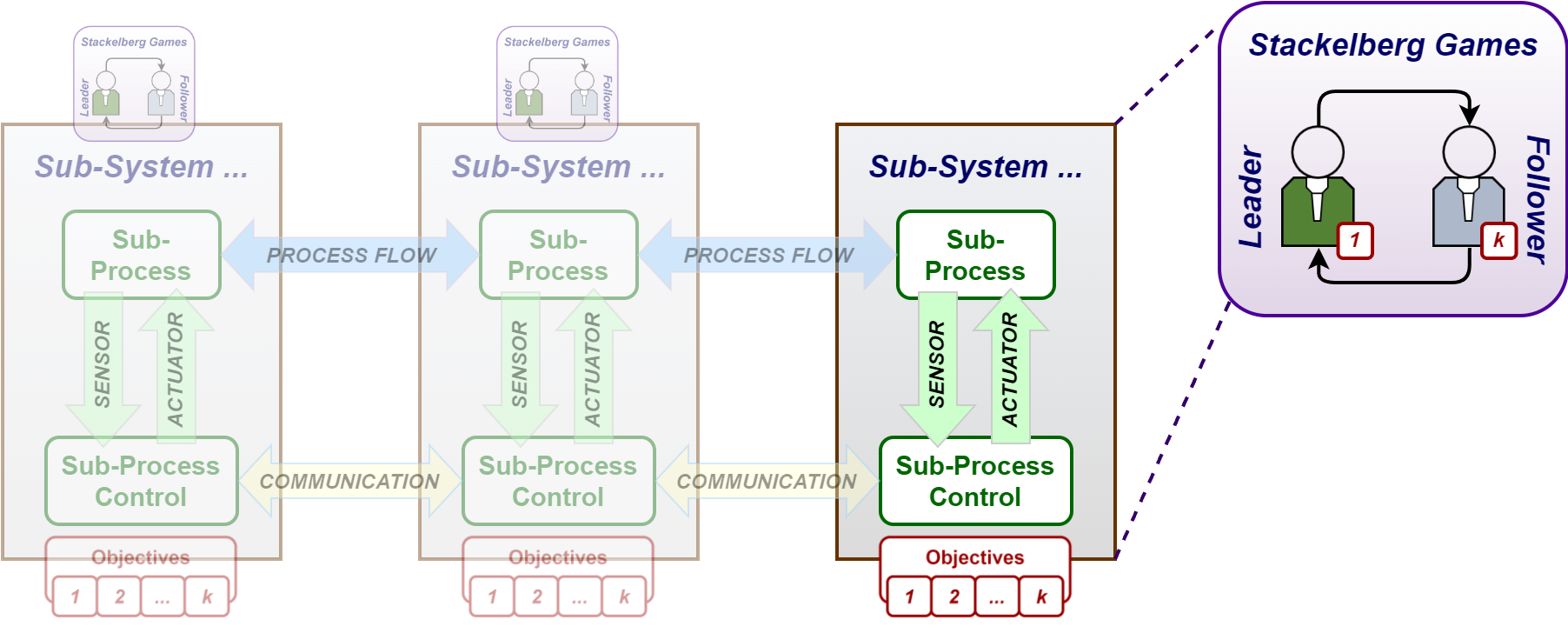}
	\caption{DS2-SbPG on a modular manufacturing unit.}
	\label{fig:modular}
\end{figure}

We propose incorporating the Stackelberg strategy as an integral component for each player $i$ in an SbPG. Using SbPG as the foundational game structure in DS2-SbPG ensures the applicability of this approach to production chains. Fig.~\ref{fig:modular} illustrates the structure of the distributed leader-follower game within a modular manufacturing unit. We design this game structure as DS2-SbPG that contains a single-leader-follower objective, defined as follows:
\begin{definition}\label{Def:DS2_SbPG}
    A game in DS2-SbPG is defined as $\Gamma(\mathcal{N}, A, S, P, \{U_i, L_i, F_i\}, \phi)$, where $\{a_1, a_2,\ldots,a_N\} \subseteq A$. Each player $i$ consists of a leader $L_i$ and a follower $F_i$, each with individual actions defined as follows:
    \begin{equation} \label{Eq:action_leader}
        a_L^i = \pi_L^i(s_i),
    \end{equation}
    \begin{equation} \label{Eq:action_follower}
        a_F^i = \pi_F^i(s_i, a_L^i),
    \end{equation}
    where $a_L^i, a_F^i \in A_i$ and $\pi_L^i, \pi_F^i$ are the governing policies of the leader and follower. An action $a_i \in A_i$ represents the coalition decision of $L_i$ and $F_i$, expressed as $a_i = a_L^i \times a_F^i$. Correspondingly, each role has its own utility function, which transforms the overall utility function $U_i = U_L^i \times U_F^i$.
\end{definition}

Each leader-follower objective pair is dedicated to optimizing a specific multi-objective function within a player $i$, with each leader-follower objective focusing on an individual objective function while preserving the convergence properties of both Stackelberg and potential games. Such convergences can be achieved through the proper selection of the governing policies $\pi_L^i$ and $\pi_F^i$, which are detailed later in Sec.~\ref{sec:ds2sbpg_3}. In DS2-SbPG, leaders hold the strategic advantage of initiating decisions before followers react.


\subsection{Stack DS2-SbPG for Multi-Leader-Follower Objective}
\label{sec:ds2sbpg_2}

\begin{figure}[t]
    \centering
    \includegraphics[width=1.0\linewidth,keepaspectratio]{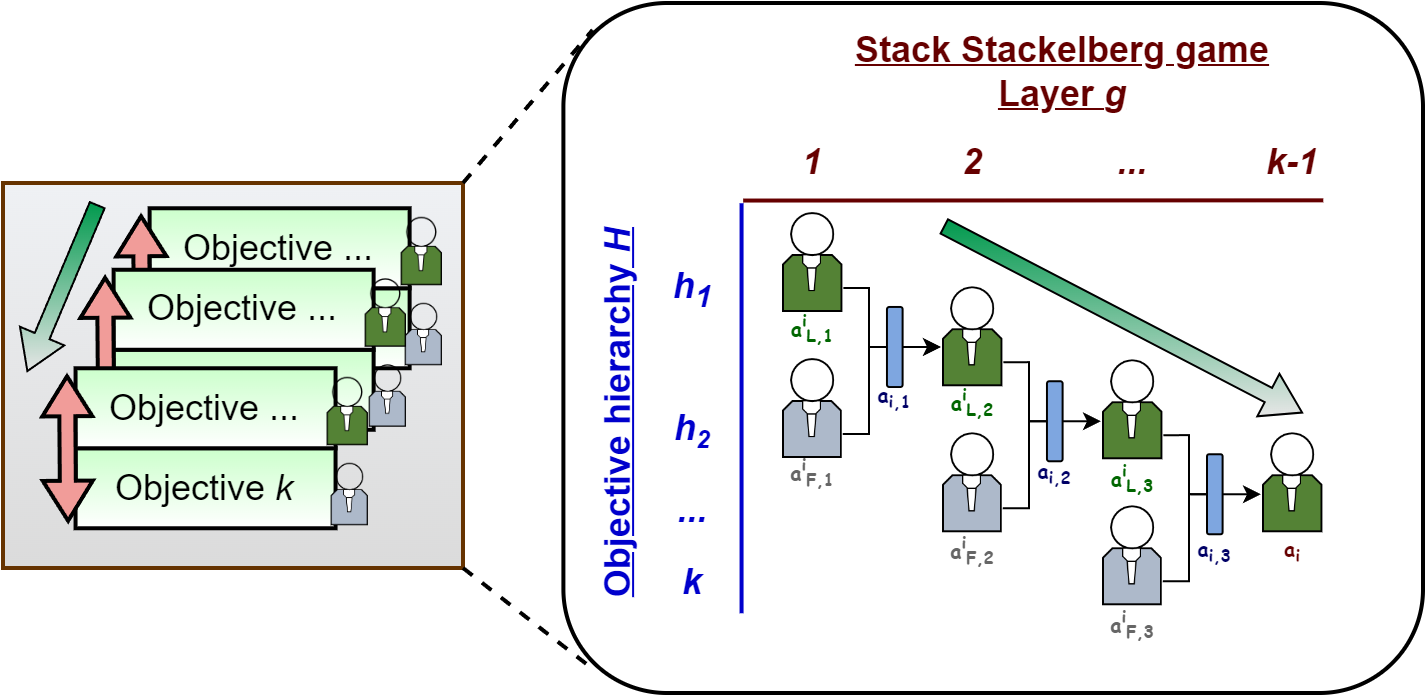}
    \caption{An illustration of the relation between the objective hierarchy $H_i$ and the stack Stackelberg games $g_i$.}
    \label{fig:stack_sg}
\end{figure}

In the second variant of DS2-SbPG, we extend the approach to accommodate strategies for multi-leader-follower objective scenarios, referred to as Stack DS2-SbPG. This variant is designed for multi-objective problems with more than two potentially conflicting objectives. In this game, each player $i$ is no longer restricted to a single leader $L_i$ and a single follower $F_i$. Instead, we introduce a hierarchical order of the roles $H_i$ based on the priority of the objectives, as illustrated in Fig.~\ref{fig:stack_sg}. Here is the formal definition of Stack DS2-SbPG:
\begin{definition}\label{def:stack}
    A Stack DS2-SbPG is defined as $\Gamma(\mathcal{N}, A, S, P, \{U_i, H_i, g_i\}, \phi)$, where $H_i$ denotes the objective hierarchy and $g_i$ represents stacked Stackelberg games in player $i$. Here, $H_i=(h_1^i,h_2^i,\ldots,h_k^i)$ is a permutation of the indices $\{1,2,\ldots,k\}$, which indicates the order in the stacked games $g_i = (g_1^i,g_2^i,\ldots,g_{k-1}^i)$. In each $g_{z}^i$, two players involved in $h_{z}^i, h_{z+1}^i$ play a leader-follower game as $L_{z}^i$ and $F_{z}^i$. The leader $L_{z}^i$ first selects an action $a_{L,z}^i(s_i) \in A_i$, see Eq.~\eqref{Eq:action_leader}, while the follower $F_{z}^i$ responds by selecting an action $a_{F,z}^i(s_i, a_{L,z}^i) \in A_i$, see Eq.~\eqref{Eq:action_follower}. The action $a_{i,z} \in A_i$ represents the coalition strategy between the leader and follower, $a_{i,z} = a_{L,z}^i \times a_{F,z}^i$. $a_{i,z}$ can then serve as the leader's action in the subsequent stacked game $a_{L,z+1}$, if $z \neq k-1$, or as the final action $a_i$, if $z = k-1$.
\end{definition}

Based on its definition, Stack DS2-SbPG involves multiple leaders and followers engaging in hierarchical Stackelberg games $g$, with each role corresponding to a different objective. For instance, if there are five objectives, $k=5$, for player $i$, there are five distinct roles ordered as $h^i_1,\ldots,h^i_5$, participating in four stacked Stackelberg games. Additionally, the order of $H_i$ indicates the importance of objectives, arranged from most to least important. The advantage of utilising multiple leader-follower pairs is the ability to accommodate numerous objectives independently, without grouping. These objectives can be arranged based on their relative importance, which allows for a comprehensive optimization approach.

\subsection{Learning Algorithm}
\label{sec:ds2sbpg_3}

After establishing the game structure, we proceed to select the learning algorithm for training the policies $\pi_L^i, \pi_F^i$ of each player $i$. Initially proposed in~\cite{Schwung2020}, we introduced the best response learning within SbPGs, which uses ad-hoc random uniform sampling processes during the learning process. With random sampling, the learning process for best response learning becomes unpredictable and unstable, as we cannot control the direction of the learning. Subsequently, we enhanced this approach to gradient-based learning~\cite{Yuwono2024}, which guides the learning direction and facilitates smoother convergence towards global optima compared to random sampling.

Both methods are developed only for simultaneous games. In contrast, DS2-SbPG involves at least one leader-follower game for each player $i$ within a potential game. In this setup, the follower can play the best-response learning, which in turn affects the leader's decision-making. Since the leader must anticipate the follower's best response, gradient-based learning is more appropriate for the leader. In the following, we provide more details on the learning algorithm for optimizing multi-objectives in DS2-SbPG without using the weighted function in Eq.~\eqref{Eq:util_sbpg}. Particularly, we have to define a representation of the player's policy, define the learning update rule for the Stackelberg Game and derive an approximate gradient descent algorithm to handle the data-based structure of the game.

\subsubsection{Policy Representation using Performance Maps}

\begin{figure}[t]
 \centering
 \includegraphics[width=1.0\columnwidth,keepaspectratio]{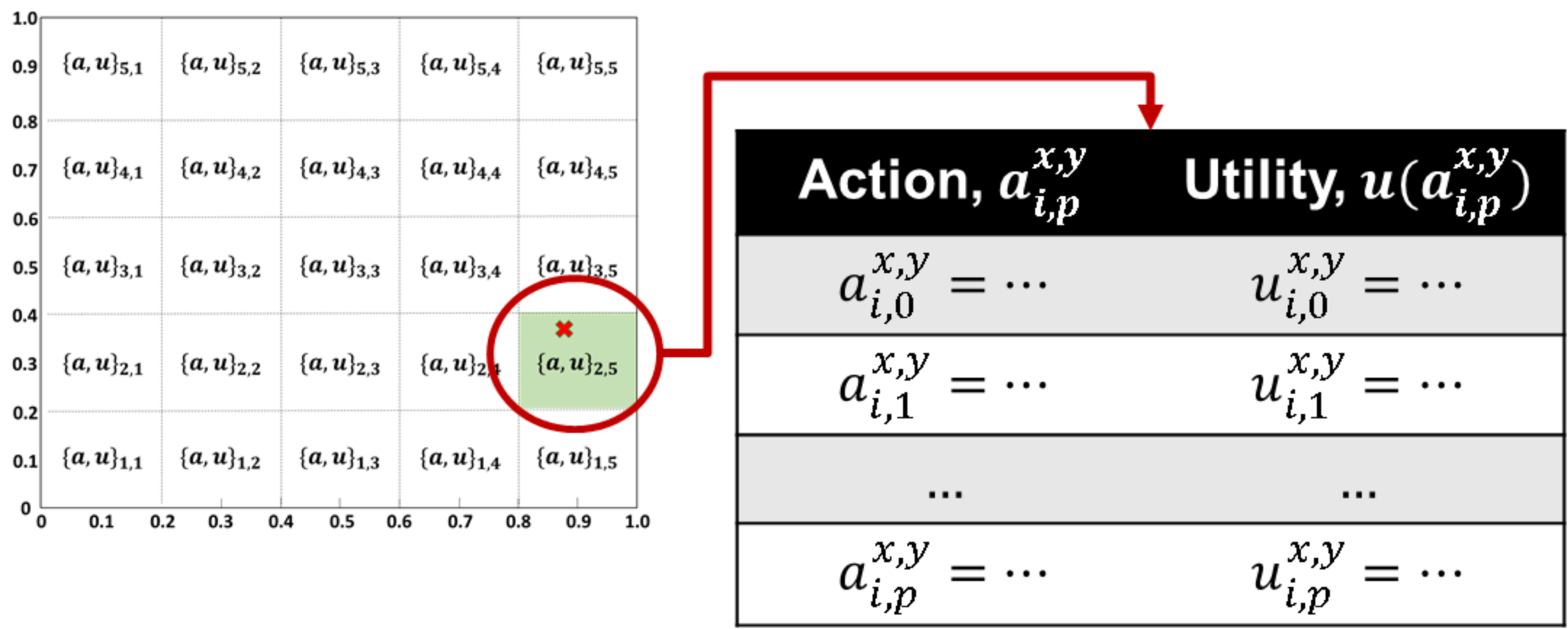}
\caption{An illustration of a 5 x 5 performance map within a 2D-state space in SbPGs~\cite{Yuwono2024}.}
\label{fig:per_maps}
\end{figure}

We first consider the representation of the policy of each player $i$ which needs to store the learned knowledge across states and actions. In SbPGs, we discretize the state space into equidistant support vectors, $l=1,\ldots,p$ which store the best-explored actions and their corresponding utility values for each state combination, alongside a stack of selected actions and their utilities for each data point across various state combinations, as proposed in~\cite{Yuwono2024}. The performance map of a player $i$ is illustrated in Fig.~\ref{fig:per_maps} for a system with two states $x$ and $y$. 

Each player $i$ selects an action $a_{i,t+1}$ by globally interpolating its map based on its current state $s_{i,t}$~\cite{Schwung2020}. Global interpolation is handled as follows:
\begin{equation} 
    w_i^{\overline{s^0s^l}} = \dfrac{1}{{(d_i^{s^0s^l})}^2 + \gamma},
\end{equation}
\begin{equation} 
    a_{i} = \sum_{l} \dfrac{w_i^{\overline{s^0s^l}}}{\sum_{q}w_i^{s^0s^l}} \cdot a_i^l,
\end{equation}
where $s^0$ denotes the current state, $s^l$ represents the state associated to the $l$-th support vector, $d_i^{s^0s^l}$ calculates the absolute distance between $s^0$ and $s^l$, $w_i^{\overline{s^0s^m}}$ calculates the weighted value, and $\gamma$ is a smoothing parameter.

In DS2-SbPG, however, we generally require maps for each leader and follower objective representing the policy associated with these objectives. A key distinction arises in the performance maps for followers, where the introduction of additional inputs as explained in Eq.~\eqref{Eq:action_follower} requires one of two potential configurations for each follower's performance map: augmentation with additional dimensions or stacking. After evaluation, the first configuration may limit players from fully exploring the state space. Even if achievable, such exploration would be time-consuming, resulting in certain grid cells remaining unexplored, thus affecting the accuracy of interpolation calculations.

\begin{figure}[t]
	\centering
	\includegraphics[width=0.7\linewidth,keepaspectratio]{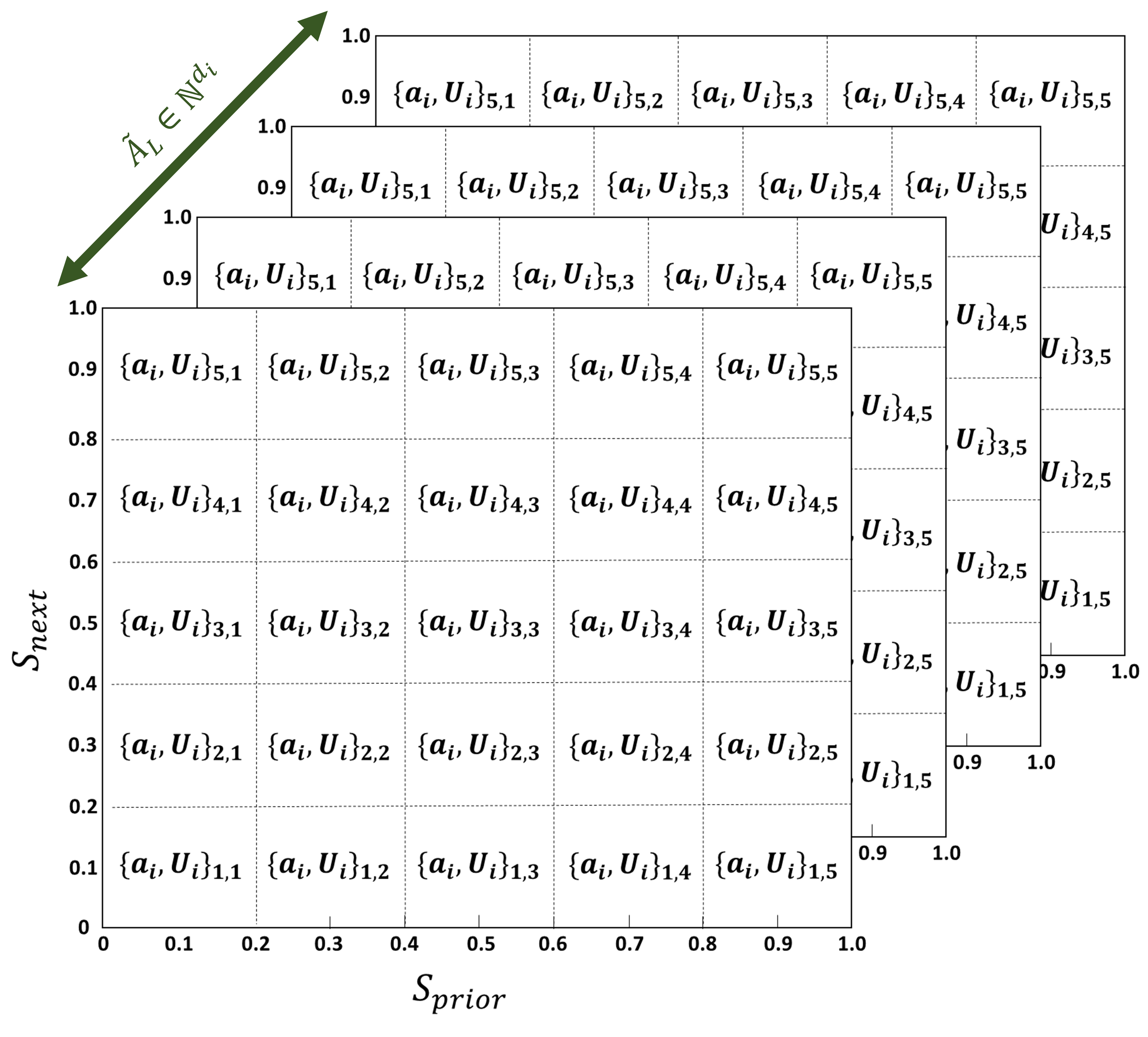}
	\caption{An updated structure of a stacking performance map of a follower in DS2-SbPG.}
	\label{fig:maps_sg}
\end{figure}

Therefore, the stacked performance map approach is preferred, as depicted in Fig.~\ref{fig:maps_sg}, primarily because it requires less exploration. When the follower interpolates its map, it accesses a single layer of the performance map based on the selected leader's actions rather than interpolating the entire map with significantly higher dimensions.

\subsubsection{Stackelberg Game-based Gradient Update Law}

We now turn our attention to the learning law used to update the above performance, i.e. to update the action values in the support vectors. Thereby, the leader and follower execute gradient-based learning differently due to the Stackelberg rule, where the leader, as the first mover, must predict the follower's best response. We begin by deriving the Stackelberg rule for the leader $L_i$. Each action in the performance map's state combinations $a_{L,i,p}^{l}$ is treated as weights to be optimized with respect to its utility function $u_L^i$ and the follower's utility function $u_F^i$, which employs deterministic learning methods, as follows:
\begin{equation}\label{eq:action_gradientebased}
    a_{L,i,p+1}^{l} = a_{L,i,p}^{l}+\alpha \cdot \omega_{L,i},
\end{equation}
where $\omega_{L,i}$ represents the vector field of the learning gradient. To compute $\omega_{L,i}$, we adopt the Stackelberg learning dynamics proposed in~\cite{Fiez2020}, which has been demonstrated to converge in both deterministic and stochastic learning scenarios, as follows:
\begin{equation}\label{eq:gradient_leaders}
    \omega_{L,i} = \frac{\partial \hat{u}_{L,i}}{\partial a_{L,i}} - \left( \frac{\partial^2 \hat{u}_{F,i}}{\partial a_{F,i} \partial a_{L,i}} \right)^T \left( \frac{\partial^2 \hat{u}_{F,i}}{\partial {a_{F,i}}^2} \right)^{-1} \frac{\partial \hat{u}_{L,i}}{\partial a_{F,i}},
\end{equation}
where $\hat{u}_{L,i}, \hat{u}_{F,i}$ describe the approximation functions for $u_{L,i}, u_{F,i}$, respectively.

As the follower plays the best response to the leader, the Stackelberg rule for a follower $F_i$ is determined as
\begin{equation}\label{eq:action_grad}
    a_{F,i,p+1}^{l} = a_{F,i,p}^{l} + \alpha \cdot \frac{\partial \hat{u}_{F,i}}{\partial a_{F,i}} + \gamma_{ou},
\end{equation}
in which $\gamma_{ou}$ indicates an optional Ornstein-Uhlenbeck (OU) noise output for exploration. As the gradient field of the follower is identical to the gradient field of SbPG, we can apply the gradient-based learning methods as in~\cite{Yuwono2024}.

\begin{bemerk}\label{bem:gradient}
The main difference between vanilla SbPG and DS2-SbPG lies in the gradient fields resulting from the different game structures. While in vanilla SbPG, the gradient field is just a weighted sum of the individual gradients $\omega_{L} \frac{\partial \hat{u}_{L,i}}{\partial a_{L,i}}  + \omega_F \frac{\partial \hat{u}_{F,i}}{\partial a_{F,i}}$, the gradient fields of DS2-SbPG in Eq.~\eqref{eq:gradient_leaders}-\eqref{eq:action_grad} are considerably different providing a superior learning performance as shown in the experiments.
\end{bemerk}

\subsubsection{Approximate Gradient Descent for Stackelberg Games}

We need to approximate the utility functions of the leader and follower because, in real production scenarios, the players in DS2-SbPG often do not have access to these utility functions as part of the system. Without this information, it is impossible to direct the learning gradient. Hence, an approximation is necessary and computed using polynomial regression because it suits continuous gradient update, as follows:
\begin{equation} 
\begin{split}
    \hat{U}_{r,i}(a_{L,i},a_{F,i}) = & \beta_0 + \beta_1 a_{L,i} + \beta_2 a_{F,i} + \beta_3 {a_{L,i}}^2 + \beta_4 {a_{F,i}}^2 \\ & + \beta_5 a_{L,i} a_{F,i} + \ldots + \beta_{n+2} {a_{L,i}}^n a_{F,i} 
\end{split}
\end{equation}
where $r \in \{L,F\}$, $n$ is the degree of the polynomial regression, and $\beta = (\beta_0, \beta_1, \ldots, \beta_{n+2})$ represents the coefficients determined using the normal equation.

\subsubsection{Learning Algorithm for DS2-SbPG}
\label{sec:ds2sbpg_5_1}

We assume $t$ is the time step of the system, and each player $i$ must update its action $a_{i,t}$ at each time step. First, at each $t$, each player $i$ obtains the current states $s_{i,t}$ of the environment. Then, the leader $L_i$ selects an action $a_L^{i,t}(s_{i,t})$ based on the related state. Following this, the follower responds with an action $a_F^{i,t}(s_{i,t},a_L^{i,t})$ based on the related states and the leader's action. Then, the set of players' actions $A_t$ is sent to the environment, which then reacts to the given action by updating the state from $S_{t+1} \leftarrow S_t$. The environment also calculates the utility value for each role $U_L^{i,t}, U_F^{i,t}$. This process then repeats from the first step, $t \leftarrow t+1$.

Algorithm~\ref{alg:dssbpg_1} provides the pseudocode for DS2-SbPG. During the training process of Stackelberg strategies, in each time $t$, followers encounter multi-step optimization of their policy $\pi_F^i$ while maintaining the leaders' strategies $a_L^{i,t}$ constant, as proposed in~\cite{Fiez2020}. This approach allows followers' strategies to converge closer before optimizing and updating the leaders' strategies.

\begin{algorithm}
\caption{Basic of Stackelberg Learning Dynamics in DS2-SbPG.}\label{alg:dssbpg_1}
\KwData{$T_{max}, \alpha, S_0, A_0$}
\For{$t=0,1,\ldots,T_{max}$}{
    \For{each player $i$}{
        obtain $l,p$ according to $s_{i,t}$;\\
        $a_{L,i,p}^{l} \leftarrow \pi_L^i(s_{i,t})$;\\
        $a_{F,i,p}^{l} \leftarrow \pi_F^i(s_{i,t}, a_{L,i,p}^{l})$;\\
        $\hat{U}_{r,i}(a_{L,i},a_{F,i}) \leftarrow \beta_0 + \beta_1 a_{L,i} + \beta_2 a_{F,i} + \beta_3 {a_{L,i}}^2 + \beta_4 {a_{F,i}}^2 + \beta_5 a_{L,i} a_{F,i} + \ldots + \beta_{n+2} {a_{L,i}}^n a_{F,i}$, $r \in \{L,F\}$;\\
        $\omega_{L,i} \leftarrow \frac{\partial \hat{U}_{L,i}}{\partial a_{L,i}} - \left( \frac{\partial^2 \hat{U}_{F,i}}{\partial a_{F,i} \partial a_{L,i}} \right)^T \cdot \left( \frac{\partial^2 \hat{U}_{F,i}}{\partial {a_{F,i}}^2} \right)^{-1} \cdot \frac{\partial \hat{U}_{L,i}}{\partial a_{F,i}}$;\\
        $\omega_{F,i} \leftarrow \frac{\partial \hat{U}_{F,i}}{\partial a_{F,i}}$;\\
        $a_{L,i,p+1}^{l} \leftarrow a_{L,i,p}^{l}+\alpha \cdot \omega_{L,i}$;\\
        $a_{F,i,p+1}^{l} \leftarrow a_{F,i,p}^{l} + \alpha \cdot \omega_{F,i} + \gamma_{ou}$;\\
        $a_L^{i,t} \leftarrow a_{L,i,p}^{l}$;\\
        $a_F^{i,t} \leftarrow a_{F,i,p}^{l}$;\\
        $a_{i,t} = a_L^{i,t} \times a_F^{i,t}$;\\
    }
    $S_{t+1} \leftarrow S_t$;\\
    calculate $U_L^{i,t}, U_F^{i,t}$;\\
}
\end{algorithm}

\subsubsection{Learning algorithm of Stack DS2-SbPG}
\label{sec:ds2sbpg_5_2}

We assume $t$ is the time step of the system, and each player $i$ must update its action $a_{i,t}$ at each time step. First, at each $t$, each player $i$ obtains the current states $s_{i,t}$ of the environment. In each $z$-Stackelberg game, $g^i_z$ is triggered. In $g^i_z$, we focus on the two objectives $h_z^i$ and $h_{z+1}^i$ in the hierarchy $H_i$. If $z=1$, the leader $L_z^i$ acts first by selecting an action $a_{L,z}^{i,t}(s_{i,t}) \in {A}_i$. Then, the follower $F_z^i$ reacts by selecting an action $a_{F,z}^{i,t}(s_{i,t}, a_{L,z}^{i,t}) \in {A}_i$. We proceed to determine the coalition strategy $a_{i,t,z} = a_L^{i,t,z} \times a_F^{i,t,z}$. 

If $z\neq k-1$, we proceed to initiate another Stackelberg game by setting $z=z+1$, which incorporates the coalition strategy from the preceding game as the leader's strategy $a_L^{i,t,z} \leftarrow a_{i,t,z-1}$. Then, we calculate $a_{F,z}^{i,t}(s_{i,t}, a_{L,z}^{i,t})$ and $a_{i,t,z} = a_L^{i,t,z} \times a_F^{i,t,z}$. This iterative process continues until reaching the final game, $z=k-1$, where the final coalition strategy $a_i$ is achieved by setting $a_{i,t} \leftarrow a_{i,t,z}$. The set of players' actions $A_t$ is then sent to the environment, which reacts to the given action by updating the state from $S_{t+1} \leftarrow S_t$. The environment also calculates the utility value for each role $U_L^{i,t,g}, U_F^{i,t,g}$. This process then repeats from the first step, $t \leftarrow t+1$. 

Algorithm~\ref{alg:dssbpg_2} provides the pseudocode for Stack DS2-SbPG. It is important to note that multi-step optimization of the followers' policy remains valid in Stack DS2-SbPG.

\begin{algorithm}
\caption{Basic of Stackelberg Learning Dynamics in Stack DS2-SbPG.}\label{alg:dssbpg_2}
\KwData{$T_{max}, \alpha, S_0, A_0, H_i, k$}
\For{$t=0,1,\ldots,T_{max}$}{
    \For{each player $i$}{
        obtain $l,p$ according to $s_{i,t}$;\\
        \For{$z=1,2,\ldots,k-1$}{
            \eIf{z=1}{
                $L_i \leftarrow L_z^i$;\\
                $a_{L,i,p}^{l} \leftarrow \pi_L^i(s_{i,t})$;\\
            }
            {
                $a_L^{i,t,z} \leftarrow a_{i,t,z-1}$;\\
            }
            $F_i \leftarrow F_z^i$;\\
            $a_{F,i,p}^{l} \leftarrow \pi_F^i(s_{i,t}, a_L^{i,t,z})$;\\
            $\hat{U}_{r,i,p}(a_{L,i},a_{F,i}) \leftarrow \beta_0 + \beta_1 a_{L,i,p} + \beta_2 a_{F,i,p} + \beta_3 {a_{L,i,p}}^2 + \beta_4 {a_{F,i,p}}^2 + \beta_5 a_{L,i,p} a_{F,i,p} + \ldots + \beta_{n+2} {a_{L,i,p}}^n a_{F,i,p}$, $r \in \{L,F\}$;\\
            \If{z=1}{
                $\omega_{L,i} \leftarrow \frac{\partial \hat{U}_{L,i}}{\partial a_{L,i}} - \left( \frac{\partial^2 \hat{U}_{F,i}}{\partial a_{F,i} \partial a_{L,i}} \right)^T \cdot \left( \frac{\partial^2 \hat{U}_{F,i}}{\partial {a_{F,i}}^2} \right)^{-1} \cdot \frac{\partial \hat{U}_{L,i}}{\partial a_{F,i}}$;\\
                $a_{L,i,p+1}^{l} \leftarrow a_{L,i,p}^{l}+\alpha \cdot \omega_{L,i}$;\\
                $a_L^{i,t,z} \leftarrow a_{L,i,p}^{l}$;\\
            }
            $\omega_{F,i} \leftarrow \frac{\partial \hat{U}_{F,i}}{\partial a_{F,i}}$;\\
            $a_{F,i,p+1}^{l} \leftarrow a_{F,i,p}^{l} + \alpha \cdot \omega_{F,i} + \gamma_{ou}$;\\
            $a_F^{i,t,z} \leftarrow a_{F,i,p}^{l}$;\\
            $a_{i,t,z} = a_L^{i,t,z} \times a_F^{i,t,z}$;\\
        }
        $a_{i,t} \leftarrow a_{i,t,z}$;\\
    }
    $S_{t+1} \leftarrow S_t$;\\
    calculate $U_L^{i,t,g}, U_F^{i,t,g}$;\\
}
\end{algorithm}

\subsection{Proof of Convergence}
\label{sec:ds2sbpg_4}

As stated in Remark~\ref{bem:gradient}, the gradient fields of DS2-SbPG differ considerably from those of vanilla SbPG. Concurrently, the gradient fields are of utmost importance for the existence of a SbPG which in turn results in convergence proof for the learning algorithms. Hence, in this subsection, we aim to prove the existence of a SbPG under the gradient update law employed for DS2-SbPG and Stack DS2-SbPG. To this end, we first present a Lemma from ~\cite{Zazo2016} which provides sufficient conditions for the existence of a SbPG (in~\cite{Zazo2016} the game is called dynamic PG with the same properties as SbPG).


\begin{lemma}\label{lemma2} \cite{Zazo2016}
    A game $\Gamma(\mathcal{N}, A, S, P, \{U_i\}, \phi)$ is considered a SbPG if the players' utilities satisfy the following conditions:
	\begin{align}\label{eq:dpg3_1}
	\frac{{\partial}U_{i}(S^{A_i},a)}{{\partial}a_{j}}=\frac{{\partial}U_{j}(S^{A_j},a)}{{\partial}a_{i}}.
	\end{align}
	\begin{align}\label{eq:dpg3_3}
	\frac{{\partial}U_{i}(S^{A_i},a)}{{\partial}s_n}=\frac{{\partial}U_{j}(S^{A_j},a)}{{\partial}s_m},
	\end{align}
   $\forall i,j \in \mathcal{N}$, $\forall s_m \in S^{A_i}$, and $\forall s_n \in S^{A_j}$.
\end{lemma}

In~\cite{Schwung2020}, different assumptions on the definition of the utility function have been proposed which assure that the distributed optimization in modular production systems constitutes a SbPG. In what follows, we assume that the leader and the follower utility functions for DS2-SbPG fulfil Assumption 1-4 in~\cite{Schwung2020}.

We can now state the following theorem:
\begin{theorem}\label{lemma1}
    Given Assumption 1-4 in~\cite{Schwung2020}, the DS2-SbPG $\Gamma(\mathcal{N}, A, S, P, \{U_i, L_i, F_i\}, \phi)$ constitute an SbPG.
\end{theorem}

\begin{IEEEproof}\label{proof:leader}
    We first note that the follower objective update has no impact on the SbPG structure of different learning agents as the follower simply plays the best response to the leader objective. Hence, we solely have to consider the gradient update of the leader objective in Eq.~\eqref{eq:gradient_leaders}. Furthermore, Condition~\eqref{eq:dpg3_3} for vanilla SbPG contains those for DS2-SbPG. Hence, as proven in~\cite{Schwung2020}, the Condition~\eqref{eq:dpg3_3} is fulfilled if the Assumption 1-4 hold. Consequently, considering Eq.~\eqref{eq:dpg3_1} of Lemma~\ref{lemma2} and incorporating Eq.~\eqref{eq:gradient_leaders}, we obtain the following condition: 
    \begin{align}\label{eq:dpg3}
         \begin{split}
            \resizebox{1.0\hsize}{!}{$\frac{\partial {U}_{L,i}(S^{A_i},a)}{\partial a_{L,j}} - \left( \frac{\partial^2 {U}_{F,i}(S^{A_i},a)}{\partial a_{F,j} \partial a_{L,j}} \right)^T \left( \frac{\partial^2 {U}_{F,i}(S^{A_i},a)}{\partial {a_{F,j}}^2} \right)^{-1} \frac{\partial {U}_{L,i}(S^{A_i},a)}{\partial a_{F,j}} = $} \\
            \resizebox{1.0\hsize}{!}{$ \frac{\partial {U}_{L,j}(S^{A_j},a)}{\partial a_{L,i}} - \left( \frac{\partial^2 {U}_{F,j}(S^{A_j},a)}{\partial a_{F,i} \partial a_{L,i}} \right)^T \left( \frac{\partial^2 {U}_{F,j}(S^{A_j},a)}{\partial {a_{F,i}}^2} \right)^{-1} \frac{\partial {U}_{L,j}(S^{A_j},a)}{\partial a_{F,i}}, $}
         \end{split}
    \end{align}
    $\forall i,j \in \mathcal{N}$ and $a_i=\{a_{L,i},a_{F,i}\}$.

    Note, that as we assume, that leader and follower objectives are sums of individual objectives, the first-order partial derivatives of Eq.~\eqref{eq:dpg3} yield
    \begin{equation}
        \frac{\partial{U}_{L,i}}{\partial a_{L,j}}=\sum_k \frac{\partial u_{ik}}{\partial a_{L,j}}, \quad
        \frac{\partial{U}_{L,i}}{\partial a_{F,j}}=\sum_k \frac{\partial u_{ik}}{\partial a_{F,j}}.
    \end{equation}
    with the second-order partial derivatives being
    \begin{equation}
        \frac{\partial^2 {U}_{F,i}}{\partial a_{F,j} \partial a_{L,j}} = \sum_k \frac{\partial^2 u_{ik}}{\partial a_{F,j} \partial a_{L,j}}, \quad
        \frac{\partial^2 {U}_{F,i}}{{\partial a_{F,j}}^2} = \sum_k \frac{\partial^2 u_{ik}}{{\partial a_{F,j}}^2}.
    \end{equation}
    
    Substituting these derivatives into the condition in Eq.~\eqref{eq:dpg3}, we obtain
    \begin{align}\label{eq:dpg3_2}
         \begin{split}
            \resizebox{1.0\hsize}{!}{$\sum_k \frac{\partial u_{ik}}{\partial a_{L,j}} \!-\!\! \left( \sum_k \frac{\partial^2 u_{ik}}{\partial a_{F,j} \partial a_{L,j}} \right)^T \!\!\left( \sum_k \frac{\partial^2 u_{ik}}{{\partial a_{F,j}}^2} \right)^{-1} \sum_k \frac{\partial u_{ik}}{\partial a_{F,j}} = $} \\
            \resizebox{1.0\hsize}{!}{$\sum_k \frac{\partial u_{jk}}{\partial a_{L,i}} \!-\!\! \left( \sum_k \frac{\partial^2 u_{jk}}{\partial a_{F,i} \partial a_{L,i}} \right)^T \!\! \left( \sum_k \frac{\partial^2 u_{jk}}{{\partial a_{F,i}}^2} \right)^{-1} \sum_k \frac{\partial u_{jk}}{\partial a_{F,i}}. $}
         \end{split}
    \end{align}
    
    However, by Assumption 4 in~\cite{Schwung2020}, the local utilities depend solely on their own action $a_i$. Hence, we have
    \begin{equation}
        \frac{\partial u_{ik}}{\partial a_{L,j}} = \frac{\partial u_{ik}}{\partial a_{F,j}} = 0, \quad
        \frac{\partial^2 u_{ik}}{\partial a_{F,j} \partial a_{L,j}} = 0.
    \end{equation}
    Hence, Condition~\eqref{eq:dpg3} is fulfilled concluding the proof.
\end{IEEEproof}

In Stack DS2-SbPG, the relationship between the multi-leader and follower objectives is structured hierarchically. Initially, the leader and follower objectives play a DS2-SbPG. Subsequently, the resulting coalition strategy from this interaction assumes the role of the leader in the next level of the stacked game, where it again plays a DS2-SbPG with a new follower objective. This iterative procedure continues until all objectives are stacked resulting in a nested gradient update by iteratively applying Eq.~\eqref{eq:gradient_leaders}. For stack DS2-SbPG, we can state the following: 
\begin{theorem}\label{lemma3}
    Given Assumption 1-4 in~\cite{Schwung2020}, the stack DS2-SbPG $ \Gamma(\mathcal{N}, A, S, P, \{U_i, H_i, g_i\}, \phi)$ constitute an SbPG.
\end{theorem}
\begin{IEEEproof}\label{proof:stacked}
Follows directly by applying the proof for DS2-SbPG iteratively for every stack.
\end{IEEEproof}

\section{Experimental Results and Discussion}
\label{sec:test}

We apply the DS2-SbPG methodology to a laboratory testbed, specifically a Bulk Good Laboratory Plant (BGLP). This implementation directly compares with our prior work in~\cite{Schwung2020} with simultaneous games. This section contains an overview of the BGLP, the setup for training DS2-SbPG on the BGLP, experimental findings, and comparisons among SbPG, DS2-SbPG, and Stack DS2-SbPG.

\subsection{Testing Environment: The Bulk Good Laboratory Plant}
\label{sec:bglp}

\begin{figure}[t]
	\centering
	\includegraphics[width=1.0\linewidth,keepaspectratio]{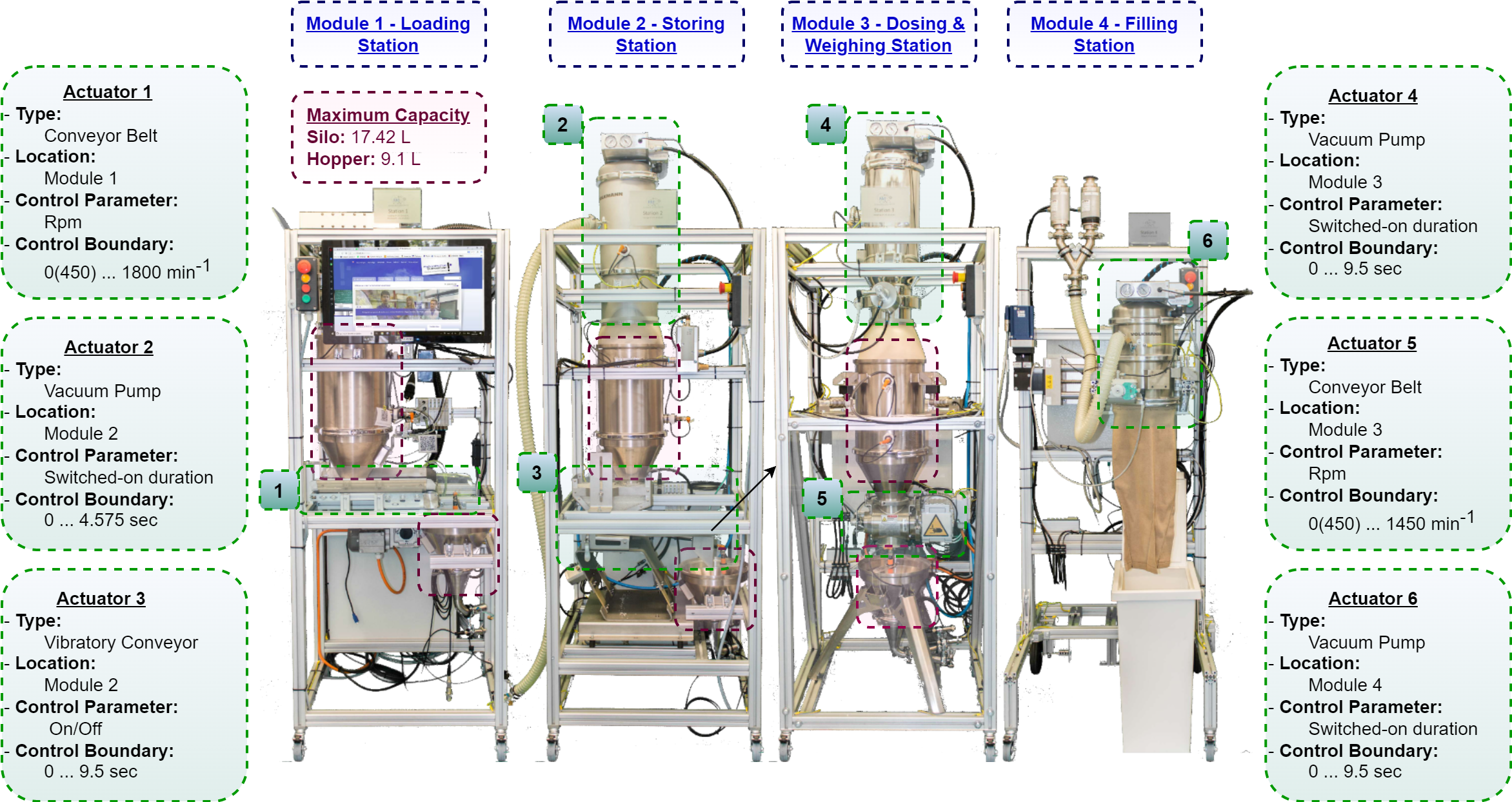}
	\caption{The Bulk Good Laboratory Plant, including the details of its modules, actuators and reservoirs.}
	\label{fig:bglp}
\end{figure}

The BGLP~\cite{Schwung2020, Yuwono2023b} is a decentralized manufacturing system with modular functionalities for bulk goods transportation. This system manages the transport of bulk goods through a network of various actuators and reservoirs, which enables transfers between four operational modules: loading, storage, weighing, and filling, as depicted in Fig.~\ref{fig:bglp}.

In Module 1, the control parameter for the conveyor belt is the motor rotation speed in rpm, similar to the rotary feeder in Module 3. The vibratory conveyor in Module 2 uses a binary value to indicate whether it is fully operational. For the vacuum pumps in Modules 2, 3, and 4, the control parameter is the turn-on duration in seconds.



The simulation of the BGLP is available in the MLPro\footnote{https://github.com/fhswf/MLPro} framework~\cite{Arend2022a}. Alternatively, the MLPro-MPPS\footnote{https://github.com/fhswf/MLPro-MPPS} framework~\cite{Yuwono2023d} provides also the simulation, which operates under a discrete event system. Those simulations are useful for training the proposed DS2-SbPG mechanism before deployment of the trained policies in the real system.

\subsection{Training Setup on the BGLP}
\label{sec:training}

In DS2-SbPG, as explained by the graph theory in Sec.~\ref{sec:prob}, each player $i$ corresponds to an individual actuator. Within the BGLP, each player $i$ has two state information, including the fill levels of the preceding and subsequent reservoirs, and at least two objectives. These objectives contain: (1) maintaining the fill level of the preceding and subsequent reservoirs to prevent bottlenecks and overflow, $u_{V}^i$ and (2) minimizing power consumption, $u_{P}^i$. Additionally, the final player is tasked with an additional objective, such as meeting production demand, $u_{D}^i$.

The objectives are evaluated at each time step $t$, with evaluation functions as outlined below:
\begin{equation} \label{Eq:obj_1}
\begin{split}
    u_{V}^i = \frac{1}{1\!+\!V_{p}^{i}} + \frac{1}{1\!+\!V_{s}^{i}}; \quad
    u_{P}^i = \frac{1}{1\!+\!P^{i}}; \quad
    u_{D}^i = \frac{1}{1\!-\!V_{D}}; \\
\end{split}
\end{equation}
where $V_{D}$ denotes the production demand fulfilment, $P^i$ the power consumption and $V_{p}^{i}, V_{s}^{i}$ represent local constraints aimed at preventing bottleneck and overflow in the prior and subsequent buffers. The production demand fulfilment $V_{D}$ is calculated as
\begin{equation}\label{Equation:4-2-1_DemandVolume}
V_{D}=\smashoperator{\int_{0}^{T_{I}}}\dot{D}_t\,dt, \quad \dot{D}_t=
\begin{dcases}
\dot{V}_{N,out}-\dot{V}_{N,in}, &\text{if}\ h^{N}=0,\\
0, &\text{otherwise,}\\
\end{dcases}
\end{equation} 
in which $\dot{V}_{N,out}$, $\dot{V}_{N,in}$ denote the outflow and inflow to the buffer, $q^{N}$ is the normalized value of the fill level, and $T_I$ represents the duration of an iteration. The demand is considered fulfilled if the fill level of the last hopper exceeds the specified production demand. Meanwhile, the mathematical expressions for $V_{p}^{i}$ and $V_{s}^{i}$ are provided as follows:
\begin{equation} \label{Equation:4-2-1_Badewanne}
V_{p}^{i}=\smashoperator{\int_{0}^{T_{I}}}\mathbbm{1}_{q^i_{p}<Q^i_{p}}(q^i_{p})\,dt, \quad V_{s}^{i}=\smashoperator{\int_{0}^{T_{I}}}\mathbbm{1}_{q^i_{s}>Q^i_{s}}(q^i_{s})\,dt, 
\end{equation}
where $Q^i_{s}$ and $Q^i_{p}$ represent the upper and lower limits of the associated fill levels. Moreover, the primary objective behind $V_{p}^{i}$ and $V_{s}^{i}$ is to minimize the risk of overflow and avoid the bottleneck, which ensures that the fill levels remain within the specified range.

\subsubsection{DS2-SbPG on the BGLP}
\label{sec:training_2}

We compute the coalition strategy between the leader and follower objectives for each player $i$ by summing them up, as follows:
\begin{equation} \label{Eq:action_coal_1}
    a_i(s_i) = a_L^i(s_i) + a_F^i(s_i,a_L^i).
\end{equation}

We note that $u_{P}^i$ is not designated as the leader objective, as it would consistently opt for the minimum action within the action space, which results in complete deactivation. Hence, we assign the bottleneck and overflow prevention utility $u_{V}^i$ and, for Player 5, the production demand fulfilment $u_{D}^i$ as leader objectives. Subsequently, the followers fine-tune the leaders' strategies to optimize power consumption $u_{P}^i$. Consequently, the leader's utility function $U_L^i$ yields
\begin{equation} \label{Eq:util_leader_1}
 U_L^i(s_i, a_i)=\left\{
    \begin{array}{ll}
    u_{V}^i + u_{D}^i \text{   , if } i=N;\\
    u_{V}^i \text{   , otherwise};\\
    \end{array}
  \right.
\end{equation}
The utility function of the follower, $U_F^i$, is formulated as follows:
\begin{equation} \label{Eq:util_follower_1}
 U_F^i(s_i, a_i)=\left\{
    \begin{array}{ll}
    \beta_L \cdot (u_{V}^i + u_{D}^i) + (1-\beta_L) \cdot u_{P}^i \text{   , if } i=N;\\
    \beta_L \cdot u_{V}^i + (1-\beta_L) \cdot u_{P}^i \text{   , otherwise}.\\
    \end{array}
  \right.
\end{equation}
We include the $u_{V}^i$ and $u_{D}^i$ objectives to the followers as this avoids action outputs to always converge to zero, where $\beta_L$ serves as the parameter regulating this trade-off. 


To accelerate the learning process, we can optionally ensure that the leader achieves their objective before starting the follower's training process. This is done by setting $U_F^i = 0 \text{, if } U_L^i < \theta_L^i$, where $\theta_L^i$ represents the threshold for evaluating the leader's success in fulfilling its objectives.

\subsubsection{Stack DS2-SbPG on the BGLP}
\label{sec:training_3}

In the Stack DS2-SbPG, we no longer subset objectives to leaders' and followers' objectives due to the stacking of Stackelberg games. We further split $u_{V}^i$ into $u_{V_p}^i$ and $u_{V_s}^i$, which refer to the preceding and subsequent reservoirs, respectively. The objective hierarchy $H^i$ follows the sequence $H_i=(u_{V_s}^i,u_{V_p}^i,u_{P}^i)$, except for the final player where the hierarchy is extended to $H_i=(u_{V_s}^i,u_{V_p}^i,u_{D}^i,u_{P}^i)$. In each game $z$, its leader and follower determine the coalition strategy by summing their strategies according to the following equation:
\begin{equation} \label{Eq:action_coal_2}
    a_{i,z}(s_i) = a_{L,z}^i(s_i) + a_{F,z}^i(s_i,a_{L,z}^i).
\end{equation}

The evaluation of the leader objective, which is characterized by its utility function $U_{L,z}^i$, as expressed by the following equation:
\begin{equation} \label{Eq:util_leader_2}
    U_{L,z}^i(s_i, a_i) = \Sigma^z_{q=1} u_{h_q^i}(s_i, a_i).
\end{equation}

Meanwhile, for the follower utility function $U_{F,z}^i$, it is expressed as follows:
\begin{equation} \label{Eq:util_follower_2}
    U_{F,z}^i(s_i, a_i) = \beta_{L,z} \cdot \Sigma^z_{q=1} u_{h_q^i}(s_i, a_i) + (1-\beta_{L,z}) \cdot u_{h_{z+1}^i}(s_i, a_i).
\end{equation}

\subsubsection{Vanilla-SbPG on the BGLP}
\label{sec:training_4}

For validation, we aim to conduct a comparative analysis with respect to the vanilla SbPG. Therefore, we apply utility functions adopted from~\cite{Schwung2020}, which are outlined as follows:
\begin{equation} \label{Equation:UtilityFunction}
 U_t^i=
    \begin{cases}
    \omega_v \cdot u_{V}^i+\omega_p \cdot u_{P}^i+\omega_d \cdot u_{D}^i & \text{if } i=N,\\
    \omega_v \cdot u_{V}^i+\omega_p \cdot u_{P}^i & \text{otherwise,}
    \end{cases}
\end{equation}
where $\omega_v, \omega_p, \omega_d$ are pre-defined weights of each objective. 

\subsection{Experimental Results}
\label{sec:res}

In this subsection, we present the outcomes and discussions deriving from our experiments. We provide comparative analyses with the Vanilla SbPG approach proposed in~\cite{Schwung2020}. Every experiment is conducted under identical settings and scenarios within the BGLP, where we conduct 9 training episodes with each episode spanning 100,000 seconds of production time and targeting a production rate of 0.125 L/s. Then, we conduct an additional testing episode utilizing the utility function outlined in Eq.~\eqref{Equation:UtilityFunction}. Furthermore, each experiment undergoes hyperparameter tuning utilizing Hyperopt~\cite{Bergstra2012} with a random grid search algorithm.

First, we execute the BGLP using Vanilla-SbPG and its parameters as proposed in~\cite{Schwung2020}. Second, we proceed with experiments utilizing DS2-SbPG for the single-leader-follower objective scenario. We maintain the performance maps discretized into 40 discrete states. Through extensive hyperparameter tuning, we determine the optimal configuration, where the number of layers of the stacked performance maps for the followers is set to 15. This results in each leader possessing a $40 \times 40$ grid of performance maps, while followers possess a $15 \times 40 \times 40$ grid of performance maps. In the optimal setting, parameters $\beta_L=0.65$, $\alpha=0.40$ and $\theta_L=2.0$, with these parameters applicable to all players. For gradient-based learning with momentum for followers, we deactivate the OU noise and set the $\alpha$ and $\beta$ parameters to 0.5 and 0.4, respectively, as proposed in~\cite{Yuwono2024}.

Third, we proceed with experiments using the Stack DS2-SbPG for the multi-leader-follower objective scenario. In this experiment, we maintain the same setting for performance maps as operated in DS2-SbPG. Following the execution of hyperparameter tuning mechanisms, the optimal settings are determined. $\alpha$, $\beta_{L,1}$, $\beta_{L,2}$, and $\beta_{L,3}$ are set to 0.50, 0.50, 0.65, and 0.75 respectively. Additionally, $\theta_L$ is disabled for all Stackelberg games, and these parameters are applied uniformly across all players. We remain the parameters of gradient-based learning with momentum for followers in~\cite{Yuwono2024}.

\begin{figure*}[t]%
    \centering
    \subfloat[\centering Production demand]{{\includegraphics[width=0.965\columnwidth]{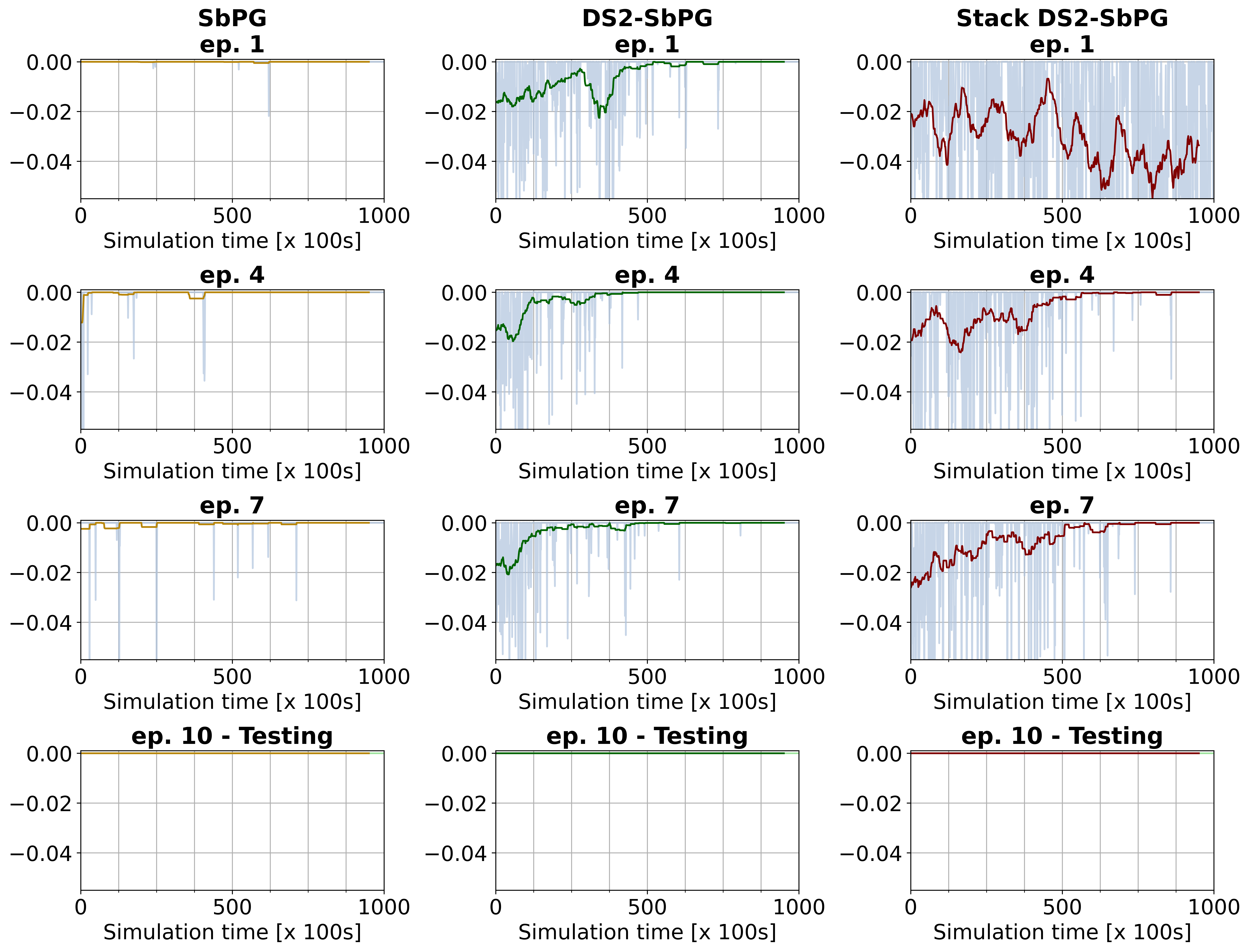} }}%
    \qquad
    \subfloat[\centering Overflow]{{\includegraphics[width=0.965\columnwidth]{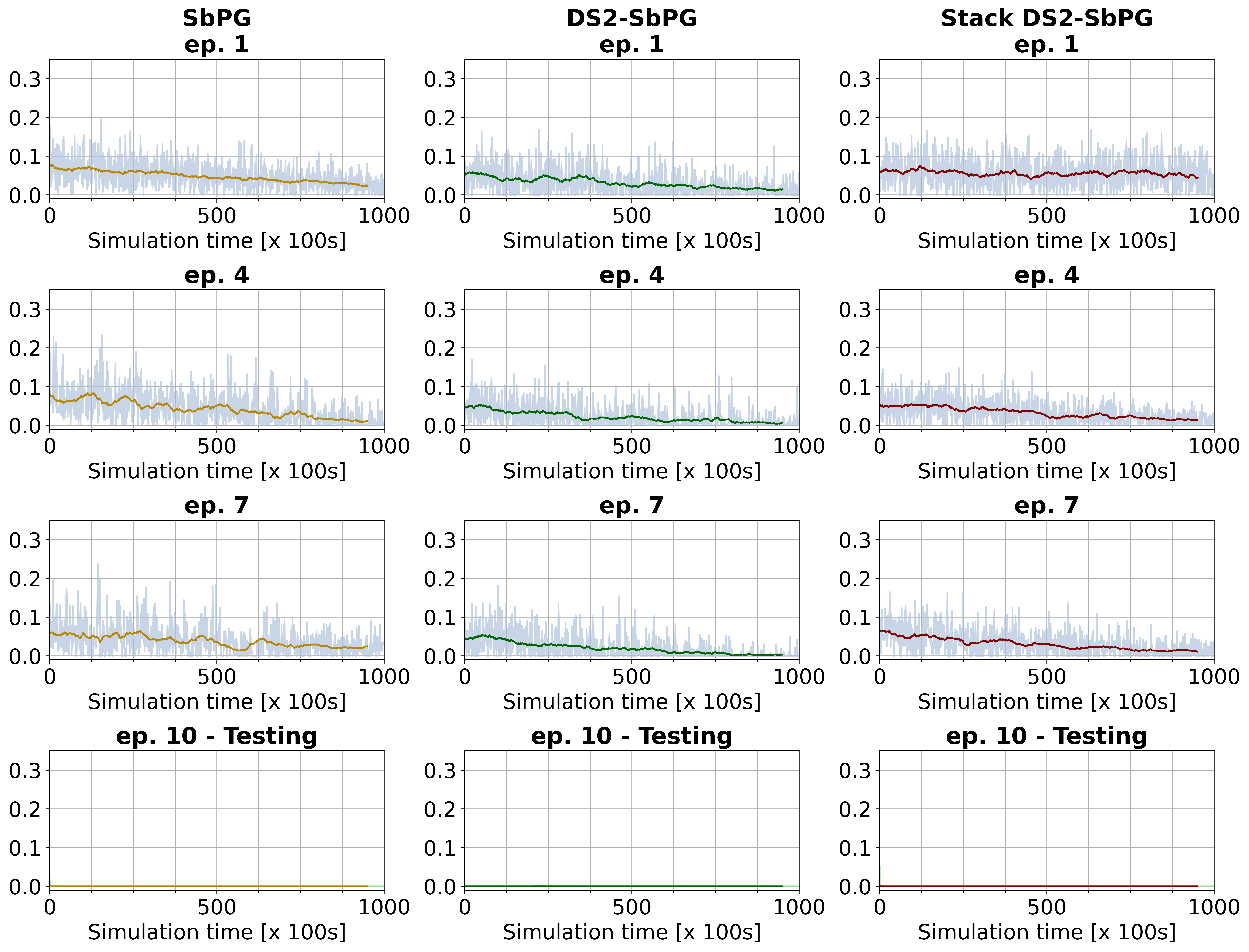} }}%
    \qquad
    \subfloat[\centering Power consumption]{{\includegraphics[width=0.965\columnwidth]{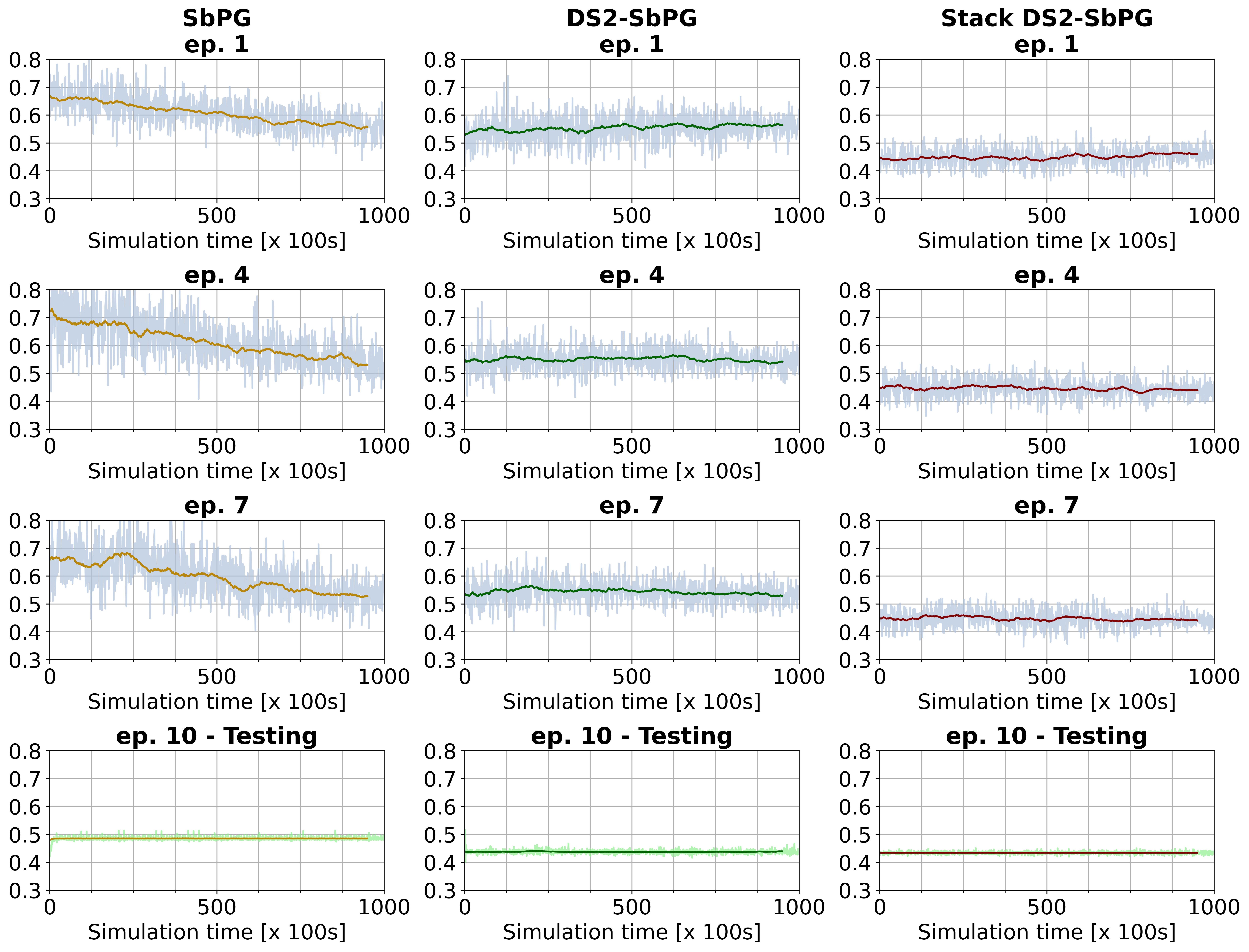} }}%
    \qquad
    \subfloat[\centering Potential value]{{\includegraphics[width=0.965\columnwidth]{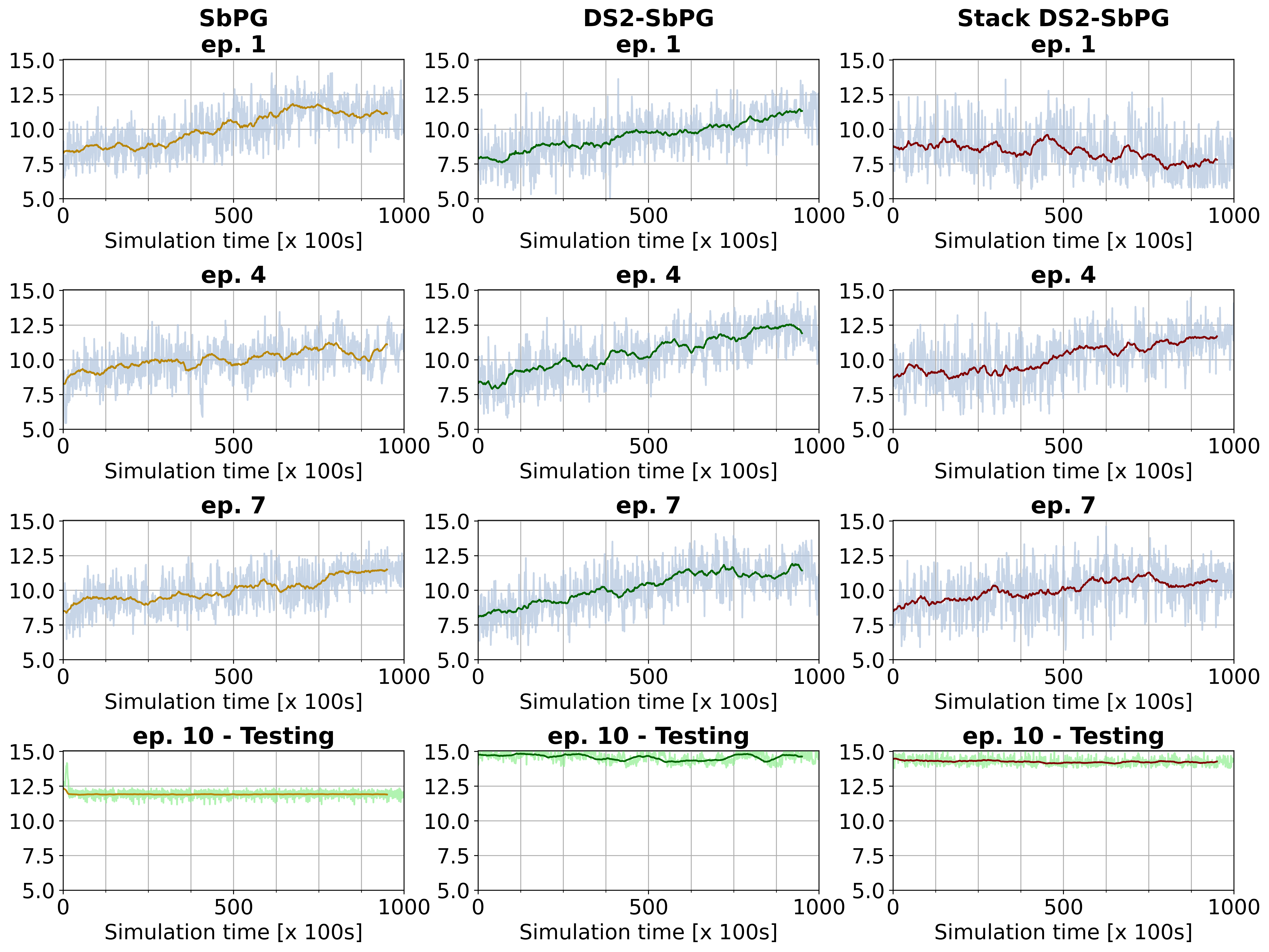} }}%
    \qquad
    \caption{Training and testing results of SbPG, DS2-SbPG, and Stack DS2-SbPG on the BGLP.}%
    \label{fig:res_all}%
\end{figure*}

Fig..~\ref{fig:res_all} shows the training and testing results of SbPG, DS2-SbPG, and Stack DS2-SbPG across episodes, which focuses on four evaluation indicators: (a) production demand fulfilment, (b) overflow, (c) power consumption, and (d) potential values. These methods display distinct learning behaviours, all demonstrating positive progress over the training period. Initially, SbPG prioritizes production demand fulfilment, possibly due to its weighting scheme, before optimizing other objectives. In contrast, DS2-SbPG and Stack DS2-SbPG optimize all objectives concurrently using the Stackelberg strategy, which leads to quicker and smoother convergence of power consumption and overflow to near-optimal levels. Stack DS2-SbPG shows a unique training pattern, initially slower but achieving optimality effectively after episode 3. Overall, all approaches exhibit continuous learning and improvement towards their objectives over time.

\begin{figure}[t]
	\centering
	\includegraphics[width=1.0\linewidth,keepaspectratio]{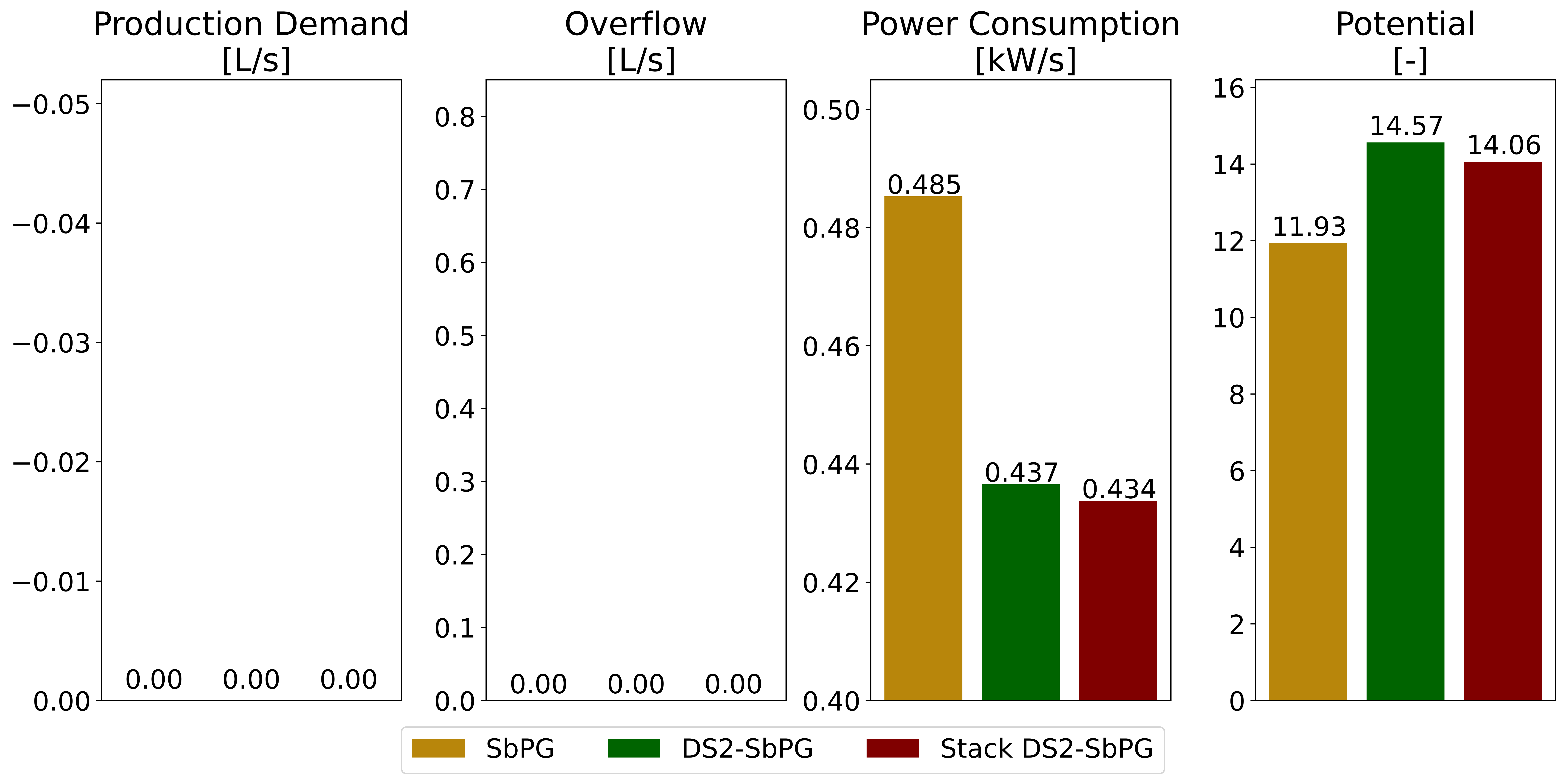}
	\caption{A comparison of the testing results between SbPG, DS2-SbPG, and Stack DS2-SbPG on the BGLP.}
	\label{fig:comparison_res}
\end{figure}

For comparisons, we present a bar graph illustrating their testing results in the BGLP, see Fig.~\ref{fig:comparison_res}. The bar graph indicates that all approaches successfully avoid overflow and fulfil the production demand. For SbPG, power consumption amounts to 0.485297 kW/s, and the average potential value reaches 11.932937. For DS2-SbPG, a notable reduction in power consumption compared to SbPG is observed, which amounts to 0.436577 kW/s and constitutes a 10.04\% decrease. Additionally, this reduction results in a higher potential value of 14.566021, which surpasses achievements in our prior research in the BGLP, as those utilising SbPGs~\cite{Schwung2020, Schwung2023, Yuwono2023b, Yuwono2024}, MARL~\cite{Schwung2021} and model predictive control~\cite{Yuwono2023a}. For Stack DS2-SbPG, the minimal power consumption of 0.433793 kW/s is achieved, which marks a 10.61\% reduction compared to SbPG. Moreover, there is a significant increase in the potential value, although it remains lower than that of DS2-SbPG.

These findings suggest that discretizing multi-objective optimization problems in Stackelberg games improves player decision-making, which simplifies the trade-off process between objectives and enables the maximization of each objective while keeping the potential games within the distributed learning. Our analysis indicates that DS2-SbPG offers a significantly simpler setup process and facilitates the determination of optimal parameters. This is because the relevant parameters are constrained within narrow spaces, as opposed to the challenge of finding weight parameters with unlimited spaces to balance the trade-offs between multiple objective functions in SbPG.

The results indicate that DS2-SbPG and Stack DS2-SbPG exhibit relatively similar performance, both outperforming SbPG significantly. However, their settings differ, where DS2-SbPG does not require the determination of subsets of optimizations for leaders and followers but necessitates the establishment of an importance hierarchy. We declare that DS2-SbPG and Stack DS2-SbPG are suited for different problem scenarios. Stack DS2-SbPG is more suitable when dealing with a higher number of objectives where no correlation exists between them, as they are entirely independent. However, it is important to highlight that DS2-SbPG demonstrates faster convergence compared to Stack DS2-SbPG, which is expected due to its smaller number of Stackelberg games involved.

\section{Conclusions}
\label{sec:conc}

We introduce a novel game structure for self-learning decentralized manufacturing systems, namely DS2-SbPG, suitable for solving multi-objective optimization problems. DS2-SbPG integrates Stackelberg strategies within each player of SbPGs while preserving a distributed approach. Unlike SbPG and other self-learning approaches in MAS that deal with determining optimal weightings for combined objective functions, DS2-SbPG simplifies the self-learning process by formulating objective trade-offs as Stackelberg games. Here, leaders prioritize crucial objectives while followers fine-tune the leaders' decisions towards other objectives. Importantly, DS2-SbPG maintains the dynamic potential game structure, which is known for its robustness and convergence. In this paper, we propose two DS2-SbPG variants, such as DS2-SbPG for single-leader-follower objective and Stack DS2-SbPG for multi-leader-follower objective. Both are implemented and evaluated in the BGLP environment, representing a smart and adaptable modular production unit, and compared against the SbPG baseline. Promising results emerge, where DS2-SbPG demonstrates a more effective setup and achieves a notable reduction in power consumption by around 10\%, and significantly enhances the potential value, indicative of optimized global objectives.

In future works, our focus will be on enhancing DS2-SbPG to be able to handle constrained optimization problems effectively. Then, we aim to elevate the performance of Best Response learning by integrating auto-concentric and gradient-based performance maps. These enhancements will be integrated into DS2-SbPG to strengthen its capabilities further. Additionally, we aspire to extend the application of this game structure by exploring its potential in diverse self-learning domains such as MARL or evolutionary algorithms.


\bibliographystyle{IEEEtran}
\bibliography{reference}

\begin{thebibliography}{10}
\providecommand{\url}[1]{#1}
\csname url@samestyle\endcsname
\providecommand{\newblock}{\relax}
\providecommand{\bibinfo}[2]{#2}
\providecommand{\BIBentrySTDinterwordspacing}{\spaceskip=0pt\relax}
\providecommand{\BIBentryALTinterwordstretchfactor}{4}
\providecommand{\BIBentryALTinterwordspacing}{\spaceskip=\fontdimen2\font plus
\BIBentryALTinterwordstretchfactor\fontdimen3\font minus \fontdimen4\font\relax}
\providecommand{\BIBforeignlanguage}[2]{{%
\expandafter\ifx\csname l@#1\endcsname\relax
\typeout{** WARNING: IEEEtran.bst: No hyphenation pattern has been}%
\typeout{** loaded for the language `#1'. Using the pattern for}%
\typeout{** the default language instead.}%
\else
\language=\csname l@#1\endcsname
\fi
#2}}
\providecommand{\BIBdecl}{\relax}
\BIBdecl

\bibitem{Fernandes20222}
M.~Fernandes, J.~M. Corchado, and G.~Marreiros, ``Machine learning techniques applied to mechanical fault diagnosis and fault prognosis in the context of real industrial manufacturing use-cases: a systematic literature review,'' \emph{Applied Intelligence}, vol.~52, no.~12, pp. 14\,246--14\,280, 2022.

\bibitem{Schwung2021}
D.~Schwung, S.~Yuwono, A.~Schwung, and S.~X. Ding, ``Decentralized learning of energy optimal production policies using plc-informed reinforcement learning,'' \emph{Computers \& Chemical Engineering}, vol. 152, p. 107382, 2021.

\bibitem{Kharitonov2022}
A.~Kharitonov, A.~Nahhas, M.~Pohl, and K.~Turowski, ``Comparative analysis of machine learning models for anomaly detection in manufacturing,'' \emph{Procedia Computer Science}, vol. 200, pp. 1288--1297, 2022.

\bibitem{Trentesaux2009}
D.~Trentesaux, ``Distributed control of production systems,'' \emph{Engineering Applications of Artificial Intelligence}, vol.~22, no.~7, pp. 971--978, 2009.

\bibitem{Bakule2008}
L.~Bakule, ``Decentralized control: An overview,'' \emph{Annual reviews in control}, vol.~32, no.~1, pp. 87--98, 2008.

\bibitem{Wooldridge2009}
M.~Wooldridge, \emph{An introduction to multiagent systems}.\hskip 1em plus 0.5em minus 0.4em\relax John wiley \& sons, 2009.

\bibitem{Yuwono2023b}
S.~Yuwono and A.~Schwung, ``Model-based learning on state-based potential games for distributed self-optimization of manufacturing systems,'' \emph{Journal of Manufacturing Systems}, vol.~71, pp. 474--493, 2023.

\bibitem{Sutton2018}
R.~S. Sutton and A.~G. Barto, \emph{Reinforcement learning: An introduction}.\hskip 1em plus 0.5em minus 0.4em\relax MIT press, 2018.

\bibitem{Okine2024}
A.~A. Okine, N.~Adam, F.~Naeem, and G.~Kaddoum, ``Multi-agent deep reinforcement learning for packet routing in tactical mobile sensor networks,'' \emph{IEEE Transactions on Network and Service Management}, 2024.

\bibitem{Orr2023}
J.~Orr and A.~Dutta, ``Multi-agent deep reinforcement learning for multi-robot applications: A survey,'' \emph{Sensors}, vol.~23, no.~7, p. 3625, 2023.

\bibitem{Chen2017}
Y.~F. Chen, M.~Liu, M.~Everett, and J.~P. How, ``Decentralized non-communicating multiagent collision avoidance with deep reinforcement learning,'' in \emph{2017 IEEE international conference on robotics and automation (ICRA)}.\hskip 1em plus 0.5em minus 0.4em\relax IEEE, 2017, pp. 285--292.

\bibitem{Cui2019}
J.~Cui, Y.~Liu, and A.~Nallanathan, ``Multi-agent reinforcement learning-based resource allocation for uav networks,'' \emph{IEEE Transactions on Wireless Communications}, vol.~19, no.~2, pp. 729--743, 2019.

\bibitem{Peng2020}
H.~Peng and X.~Shen, ``Multi-agent reinforcement learning based resource management in mec-and uav-assisted vehicular networks,'' \emph{IEEE Journal on Selected Areas in Communications}, vol.~39, no.~1, pp. 131--141, 2020.

\bibitem{Nguyen2021}
T.~T. Nguyen and V.~J. Reddi, ``Deep reinforcement learning for cyber security,'' \emph{IEEE Transactions on Neural Networks and Learning Systems}, vol.~34, no.~8, pp. 3779--3795, 2021.

\bibitem{Lowe2017}
R.~Lowe, Y.~I. Wu, A.~Tamar, J.~Harb, O.~Pieter~Abbeel, and I.~Mordatch, ``Multi-agent actor-critic for mixed cooperative-competitive environments,'' \emph{Advances in neural information processing systems}, vol.~30, 2017.

\bibitem{Rashid2020}
T.~Rashid, G.~Farquhar, B.~Peng, and S.~Whiteson, ``Weighted qmix: Expanding monotonic value function factorisation for deep multi-agent reinforcement learning,'' \emph{Advances in neural information processing systems}, vol.~33, pp. 10\,199--10\,210, 2020.

\bibitem{Foerster2018}
J.~Foerster, G.~Farquhar, T.~Afouras, N.~Nardelli, and S.~Whiteson, ``Counterfactual multi-agent policy gradients,'' in \emph{Proceedings of the AAAI conference on artificial intelligence}, vol.~32, no.~1, 2018.

\bibitem{Canese2021}
L.~Canese, G.~C. Cardarilli, L.~Di~Nunzio, R.~Fazzolari, D.~Giardino, M.~Re, and S.~Span{\`o}, ``Multi-agent reinforcement learning: A review of challenges and applications,'' \emph{Applied Sciences}, vol.~11, no.~11, p. 4948, 2021.

\bibitem{Dulac2021}
G.~Dulac-Arnold, N.~Levine, D.~J. Mankowitz, J.~Li, C.~Paduraru, S.~Gowal, and T.~Hester, ``Challenges of real-world reinforcement learning: definitions, benchmarks and analysis,'' \emph{Machine Learning}, vol. 110, no.~9, pp. 2419--2468, 2021.

\bibitem{Nowe2021}
A.~Now{\'e}, P.~Vrancx, and Y.-M. De~Hauwere, ``Game theory and multi-agent reinforcement learning,'' \emph{Reinforcement Learning: State-of-the-Art}, pp. 441--470, 2012.

\bibitem{Schwung2020}
D.~Schwung, A.~Schwung, and S.~X. Ding, ``Distributed self-optimization of modular production units: A state-based potential game approach,'' \emph{IEEE Transactions on Cybernetics}, vol.~52, no.~4, pp. 2174--2185, 2020.

\bibitem{Schwung2023}
D.~Schwung, S.~Yuwono, A.~Schwung, and S.~X. Ding, ``Plc-informed distributed game theoretic learning of energy-optimal production policies,'' \emph{IEEE Transactions on Cybernetics}, vol.~53, no.~9, pp. 5424--5435, 2023.

\bibitem{Yuwono2023c}
S.~Yuwono and A.~Schwung, ``A model-based deep learning approach for self-learning in smart production systems,'' in \emph{2023 IEEE 28th International Conference on Emerging Technologies and Factory Automation (ETFA)}.\hskip 1em plus 0.5em minus 0.4em\relax IEEE, 2023, pp. 1--8.

\bibitem{Monderer1996}
D.~Monderer and L.~S. Shapley, ``Potential games,'' \emph{Games and economic behavior}, vol.~14, no.~1, pp. 124--143, 1996.

\bibitem{Marden2012}
J.~R. Marden, ``State based potential games,'' \emph{Automatica}, vol.~48, no.~12, pp. 3075--3088, 2012.

\bibitem{Zazo2016}
S.~Zazo, S.~Valcarcel~Macua, M.~S{\'a}nchez-Fern{\'a}ndez, and J.~Zazo, ``Dynamic potential games with constraints: Fundamentals and applications in communications,'' \emph{IEEE Transactions on Signal Processing}, vol.~64, no.~14, pp. 3806--3821, 2016.

\bibitem{Bergstra2012}
J.~Bergstra and Y.~Bengio, ``Random search for hyper-parameter optimization.'' \emph{Journal of machine learning research}, vol.~13, no.~2, 2012.

\bibitem{Owen2013}
G.~Owen, \emph{Game theory}.\hskip 1em plus 0.5em minus 0.4em\relax Emerald Group Publishing, 2013.

\bibitem{Bauso2016}
D.~Bauso, \emph{Game theory with engineering applications}.\hskip 1em plus 0.5em minus 0.4em\relax SIAM, 2016.

\bibitem{Simaan1973}
M.~Simaan and J.~B. Cruz~Jr, ``On the stackelberg strategy in nonzero-sum games,'' \emph{Journal of Optimization Theory and Applications}, vol.~11, no.~5, pp. 533--555, 1973.

\bibitem{Stackelberg2010}
H.~Von~Stackelberg, \emph{Market structure and equilibrium}.\hskip 1em plus 0.5em minus 0.4em\relax Springer Science \& Business Media, 2010.

\bibitem{Fiez2020}
T.~Fiez, B.~Chasnov, and L.~Ratliff, ``Implicit learning dynamics in stackelberg games: Equilibria characterization, convergence analysis, and empirical study,'' in \emph{International Conference on Machine Learning}.\hskip 1em plus 0.5em minus 0.4em\relax PMLR, 2020, pp. 3133--3144.

\bibitem{Miettinen1999}
K.~Miettinen, \emph{Nonlinear multiobjective optimization}.\hskip 1em plus 0.5em minus 0.4em\relax Springer Science \& Business Media, 1999, vol.~12.

\bibitem{Branke2008}
J.~Branke, \emph{Multiobjective optimization: Interactive and evolutionary approaches}.\hskip 1em plus 0.5em minus 0.4em\relax Springer Science \& Business Media, 2008, vol. 5252.

\bibitem{Ngatchou2005}
P.~Ngatchou, A.~Zarei, and A.~El-Sharkawi, ``Pareto multi objective optimization,'' in \emph{Proceedings of the 13th international conference on, intelligent systems application to power systems}.\hskip 1em plus 0.5em minus 0.4em\relax IEEE, 2005, pp. 84--91.

\bibitem{Kalyanmoy2016}
K.~Deb, K.~Sindhya, and J.~Hakanen, ``Multi-objective optimization,'' in \emph{Decision sciences}.\hskip 1em plus 0.5em minus 0.4em\relax CRC Press, 2016, pp. 161--200.

\bibitem{Sharma2022}
S.~Sharma and V.~Kumar, ``A comprehensive review on multi-objective optimization techniques: Past, present and future,'' \emph{Archives of Computational Methods in Engineering}, vol.~29, no.~7, pp. 5605--5633, 2022.

\bibitem{Bertsekas2014}
D.~P. Bertsekas, \emph{Constrained optimization and Lagrange multiplier methods}.\hskip 1em plus 0.5em minus 0.4em\relax Academic press, 2014.

\bibitem{Biere2009}
A.~Biere, M.~Heule, and H.~van Maaren, \emph{Handbook of satisfiability}.\hskip 1em plus 0.5em minus 0.4em\relax IOS press, 2009, vol. 185.

\bibitem{Wenlan2019}
W.~Huang, Y.~Zhang, and L.~Li, ``Survey on multi-objective evolutionary algorithms,'' in \emph{Journal of physics: Conference series}, vol. 1288, no.~1.\hskip 1em plus 0.5em minus 0.4em\relax IOP Publishing, 2019, p. 012057.

\bibitem{Mannion2018}
P.~Mannion, S.~Devlin, J.~Duggan, and E.~Howley, ``Reward shaping for knowledge-based multi-objective multi-agent reinforcement learning,'' \emph{The Knowledge Engineering Review}, vol.~33, p. e23, 2018.

\bibitem{Yuandou2019}
Y.~Wang, H.~Liu, W.~Zheng, Y.~Xia, Y.~Li, P.~Chen, K.~Guo, and H.~Xie, ``Multi-objective workflow scheduling with deep-q-network-based multi-agent reinforcement learning,'' \emph{IEEE access}, vol.~7, pp. 39\,974--39\,982, 2019.

\bibitem{Diehl2023}
C.~Diehl, T.~Klosek, M.~Kr{\"u}ger, N.~Murzyn, and T.~Bertram, ``On a connection between differential games, optimal control, and energy-based models for multi-agent interactions,'' \emph{arXiv preprint arXiv:2308.16539}, 2023.

\bibitem{Mourtzis2013}
D.~Mourtzis and M.~Doukas, ``Decentralized manufacturing systems review: challenges and outlook,'' in \emph{Robust Manufacturing Control: Proceedings of the CIRP Sponsored Conference RoMaC 2012, Bremen, Germany, 18th-20th June 2012}.\hskip 1em plus 0.5em minus 0.4em\relax Springer, 2013, pp. 355--369.

\bibitem{Jiewu2023}
J.~Leng, Y.~Zhong, Z.~Lin, K.~Xu, D.~Mourtzis, X.~Zhou, P.~Zheng, Q.~Liu, J.~L. Zhao, and W.~Shen, ``Towards resilience in industry 5.0: A decentralized autonomous manufacturing paradigm,'' \emph{Journal of Manufacturing Systems}, vol.~71, pp. 95--114, 2023.

\bibitem{Du2024}
Y.~Du and J.-q. Li, ``A deep reinforcement learning based algorithm for a distributed precast concrete production scheduling,'' \emph{International Journal of Production Economics}, vol. 268, p. 109102, 2024.

\bibitem{Lee2024}
C.~Lee, S.~Zhang, Y.~Tsang, and J.~Huo, ``Resource allocation for distributed production networks,'' in \emph{Digital Manufacturing}.\hskip 1em plus 0.5em minus 0.4em\relax Elsevier, 2024, pp. 247--277.

\bibitem{Rokhforoz2021}
P.~Rokhforoz and O.~Fink, ``Distributed joint dynamic maintenance and production scheduling in manufacturing systems: Framework based on model predictive control and benders decomposition,'' \emph{Journal of Manufacturing Systems}, vol.~59, pp. 596--606, 2021.

\bibitem{Li2020}
T.~Li, A.~K. Sahu, A.~Talwalkar, and V.~Smith, ``Federated learning: Challenges, methods, and future directions,'' \emph{IEEE signal processing magazine}, vol.~37, no.~3, pp. 50--60, 2020.

\bibitem{Ilie2013}
S.~Ilie and C.~B{\u{a}}dic{\u{a}}, ``Multi-agent distributed framework for swarm intelligence,'' \emph{Procedia Computer Science}, vol.~18, pp. 611--620, 2013.

\bibitem{Yuwono2023a}
S.~Yuwono and A.~Schwung, ``Model predictive control with adaptive plc-based policy on low dimensional state representation for industrial applications,'' in \emph{2023 31st Mediterranean Conference on Control and Automation (MED)}.\hskip 1em plus 0.5em minus 0.4em\relax IEEE, 2023, pp. 883--889.

\bibitem{Borel1953}
E.~Borel, ``The theory of play and integral equations with skew symmetric kernels,'' \emph{Econometrica: journal of the Econometric Society}, pp. 97--100, 1953.

\bibitem{Rosenthal1973}
R.~W. Rosenthal, ``A class of games possessing pure-strategy nash equilibria,'' \emph{International Journal of Game Theory}, vol.~2, pp. 65--67, 1973.

\bibitem{Tamer2018}
T.~Ba{\c{s}}ar and G.~Zaccour, \emph{Handbook of dynamic game theory}.\hskip 1em plus 0.5em minus 0.4em\relax Springer, 2018.

\bibitem{Lim2021}
W.~Y.~B. Lim, J.~S. Ng, Z.~Xiong, J.~Jin, Y.~Zhang, D.~Niyato, C.~Leung, and C.~Miao, ``Decentralized edge intelligence: A dynamic resource allocation framework for hierarchical federated learning,'' \emph{IEEE Transactions on Parallel and Distributed Systems}, vol.~33, no.~3, pp. 536--550, 2021.

\bibitem{Zhaolong2021}
Z.~Ning, Y.~Yang, X.~Wang, L.~Guo, X.~Gao, S.~Guo, and G.~Wang, ``Dynamic computation offloading and server deployment for uav-enabled multi-access edge computing,'' \emph{IEEE Transactions on Mobile Computing}, vol.~22, no.~5, pp. 2628--2644, 2021.

\bibitem{Yan2014}
Y.~Li, L.~Xiao, J.~Liu, and Y.~Tang, ``Power control stackelberg game in cooperative anti-jamming communications,'' in \emph{The 2014 5th International Conference on Game Theory for Networks}, 2014, pp. 1--6.

\bibitem{Yuzhe2018}
Y.~Li, D.~Shi, and T.~Chen, ``False data injection attacks on networked control systems: A stackelberg game analysis,'' \emph{IEEE Transactions on Automatic Control}, vol.~63, no.~10, pp. 3503--3509, 2018.

\bibitem{Luliang2018}
L.~Jia, Y.~Xu, Y.~Sun, S.~Feng, and A.~Anpalagan, ``Stackelberg game approaches for anti-jamming defence in wireless networks,'' \emph{IEEE Wireless Communications}, vol.~25, no.~6, pp. 120--128, 2018.

\bibitem{Yamamoto2015}
K.~Yamamoto, ``A comprehensive survey of potential game approaches to wireless networks,'' \emph{IEICE Transactions on Communications}, vol.~98, no.~9, pp. 1804--1823, 2015.

\bibitem{Yuwono2024}
S.~Yuwono, M.~L{\"o}ppenberg, D.~Schwung, and A.~Schwung, ``Gradient-based learning in state-based potential games for self-learning production systems,'' 2024.

\bibitem{Arend2022a}
D.~Arend, S.~Yuwono, M.~R. Diprasetya, and A.~Schwung, ``Mlpro 1.0-standardized reinforcement learning and game theory in python,'' \emph{Machine Learning with Applications}, vol.~9, p. 100341, 2022.

\bibitem{Yuwono2023d}
S.~Yuwono, M.~L{\"o}ppenberg, D.~Arend, M.~R. Diprasetya, and A.~Schwung, ``Mlpro-mpps - a versatile and configurable production systems simulator in python,'' in \emph{2023 IEEE 2nd Industrial Electronics Society Annual On-Line Conference (ONCON)}, 2023, pp. 1--6.

\end{thebibliography}

\end{document}